\documentclass[sigplan,10pt]{acmart}
\renewcommand\footnotetextcopyrightpermission[1]{}
\usepackage[normalem]{ulem}
\usepackage{graphicx}
\usepackage{textcomp}
\usepackage{multirow}
\usepackage{colortbl}
\usepackage{xcolor}
\usepackage{verbatim}
\usepackage{appendix}
\usepackage{bbding}
\usepackage{booktabs}
\usepackage{listings}
\usepackage{subcaption}
\usepackage{epstopdf}
\usepackage{xcolor}
\usepackage{tikz}

\begin{document}

\title{Bridging the Gap Between Domain-specific Frameworks and Multiple Hardware Devices}
\author{Xu Wen}
\affiliation{%
  \institution{Institute of Computing Technology, Chinese Academy of Sciences \and University of Chinese Academy of Sciences}
  \city{Beijing}
  \country{China}
}
\email{wenxu@ict.ac.cn}

\author{Wanling Gao}
\affiliation{%
  \institution{Institute of Computing Technology, Chinese Academy of Sciences \and University of Chinese Academy of Sciences}
  \city{Beijing}
  \country{China}
}
\email{gaowanling@ict.ac.cn}

\author{Lei Wang}
\affiliation{%
  \institution{Institute of Computing Technology, Chinese Academy of Sciences \and University of Chinese Academy of Sciences}
  \city{Beijing}
  \country{China}
}
\email{wanglei_2011@ict.ac.cn}

\author{Jianfeng Zhan}
\authornote{Corresponding author.}
\affiliation{%
  \institution{Institute of Computing Technology, Chinese Academy of Sciences \and University of Chinese Academy of Sciences}
  \city{Beijing}
  \country{China}
}
\email{zhanjianfeng@ict.ac.cn}

\begin{abstract}

% - how to bridge the gap between these frameworks and the diverse array of hardware architectures and devices available

The rapid development of domain-specific frameworks has presented us with a significant challenge: The current approach of implementing solutions on a case-by-case basis incurs a theoretical complexity of O(M$\times$N), thereby increasing the cost of porting applications to different hardware platforms. 
To address these challenges, we propose a systematic methodology that effectively bridges the gap between domain-specific frameworks and multiple hardware devices, reducing porting complexity to O(M+N). The approach utilizes multi-layer abstractions.  Different domain-specific abstractions are employed to represent applications from various domains. These abstractions are then transformed into a unified abstraction, which is subsequently translated into combinations of primitive operators. Finally, these operators are mapped to multiple hardware platforms. The implemented unified framework supports deep learning, classical machine learning, and data analysis across X86, ARM, RISC-V, IoT devices, and GPU. It outperforms existing solutions like scikit-learn, hummingbird, Spark, and pandas, achieving impressive speedups: 1.1x to 3.83x on X86 servers, 1.06x to 4.33x on ARM IoT devices, 1.25x to 3.72x on RISC-V IoT devices, and 1.93x on GPU.
The source code is available at \url{https://github.com/BenchCouncil/bridger.git}.
\end{abstract}


\maketitle
\pagestyle{plain}
\section{Introduction}\label{introduction}

Different application domains exhibit distinct characteristics in terms of data and computation. In deep learning (DL), tensors are prevalent, and iterative processes driven by backpropagation are essential~\cite{werbos1990backpropagation}. Classical machine learning (CML) primarily deals with arrays, although certain algorithms can handle tables and tensors as well. CML also involves iterative processes, but they are not specifically tied to backpropagation. Data analytics (DA) mainly revolves around tables and arrays, with computations that do not necessarily require iteration.

To cater to the specific requirements of each domain, provide user-friendly interfaces, and promote code reuse, current high-level applications often rely on domain-specific frameworks. Some popular frameworks include TensorFlow~\cite{10.5555/3026877.3026899}, PyTorch~\cite{NEURIPS2019_bdbca288}, and Mxnet~\cite{chen2015mxnet} for DL; scikit-learn~\cite{10.5555/1953048.2078195}, Spark MLlib~\cite{10.5555/2946645.2946679}, and XGBoost~\cite{chen2016xgboost} for CML; and pandas~\cite{mckinney2011pandas}, Spark~\cite{zaharia2010spark}, and Dask~\cite{rocklin2015dask} for DA. These frameworks enable developers to work efficiently within their respective domains.

The rapid development of domain-specific frameworks has presented us with a significant challenge: the current approach of implementing solutions on a case-by-case basis incurs a theoretical complexity of O(M$\times$N), thereby increasing the cost of porting applications to different hardware platforms. As depicted in Fig.~\ref{fig_status}, we can observe the existence of numerous domain-specific frameworks in the domains of DL, CML, and DA. However, these frameworks often lack comprehensive support for the wide range of underlying hardware options.

Frameworks like scikit-learn and pandas, which are widely utilized in CML and DA, are primarily designed for hardware with mature software ecosystems such as X86 and ARM. Consequently, they may not possess the necessary support and optimizations for emerging hardware architectures like RISC-V, GPUs, and Internet of Things (IoT) devices.

This gap poses challenges for both software and hardware utilization. From a software perspective, the migration of applications to different hardware platforms becomes significantly challenging and costly. On the hardware side, emerging hardware devices face limitations in their development due to inadequate software ecosystems.

\begin{figure}
  \centering
  \begin{subfigure}[b]{0.49\textwidth}
    \includegraphics[width=\textwidth]{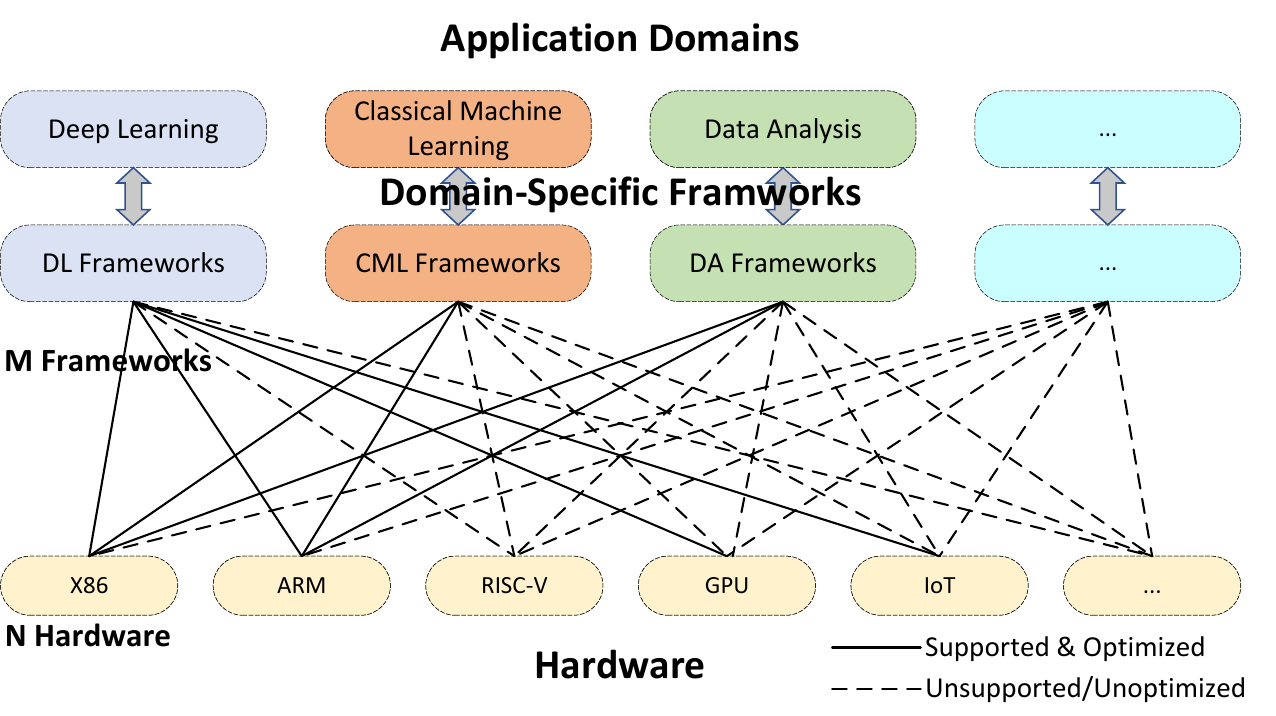} 
    \caption{Current domain-specific frameworks lack support and optimizations for diverse hardware devices. Case-by-case implementations have a complexity of O(M$\times$N).}
    \label{fig_status}
  \end{subfigure}
  \begin{subfigure}[b]{0.49\textwidth}
    \includegraphics[width=\textwidth]{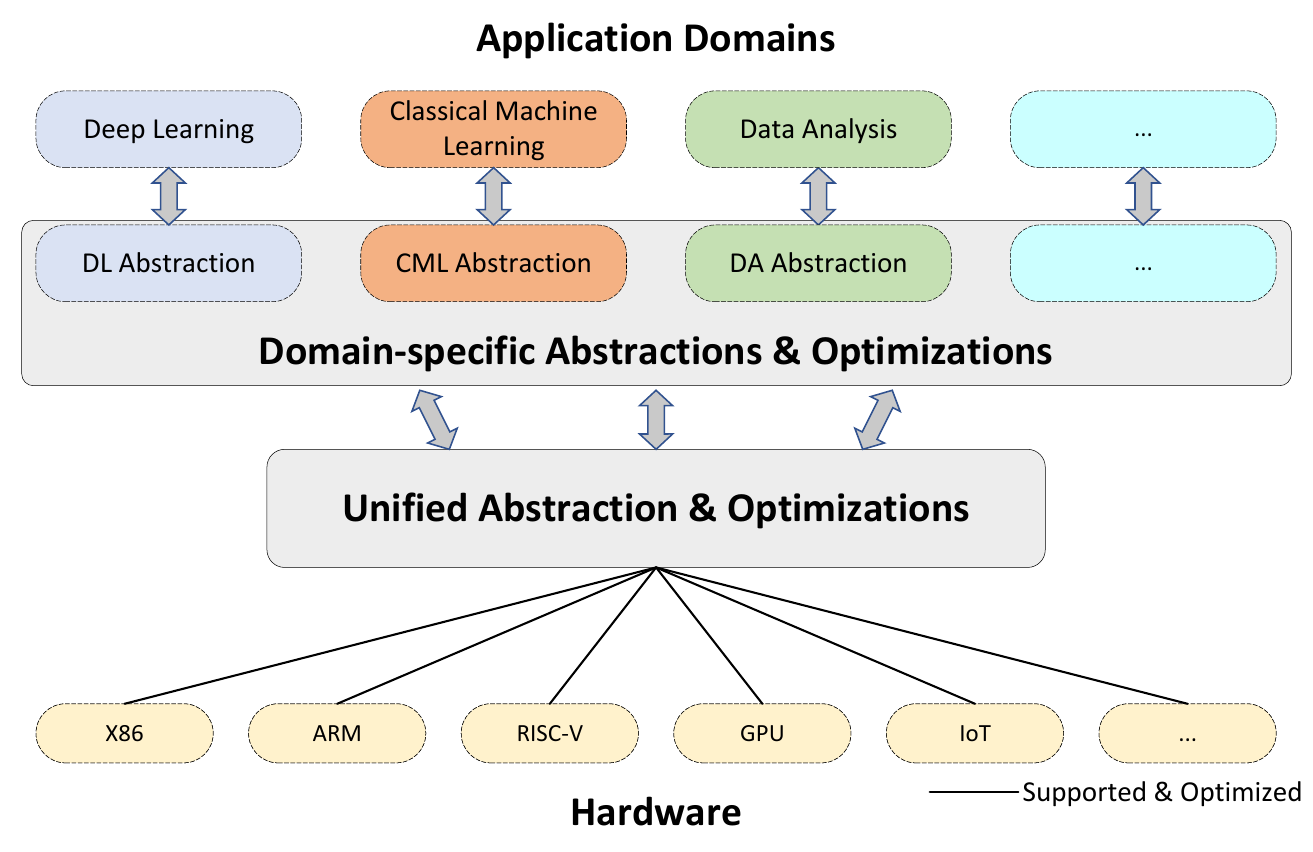}
    \caption{We use domain-specific abstractions and unified abstractions to bridge applications and hardware, reducing porting complexity from O(M$\times$N) to O(M+N).}
    \label{fig_method}
  \end{subfigure}
  \caption{We propose a method to systematically bridge the gap between domain-specific frameworks and multiple hardware devices. }
  \label{fig_motivation}
\end{figure}

Additionally, domain-specific frameworks exist in isolation, with the progress made in one application domain unable to be shared with others, resulting in a significant amount of duplicated work.
DL has garnered significant attention and investment, benefiting from relatively robust hardware support and optimizations~\cite{mohammadi2018deep,debauche2020new,vrevca2020accelerating,zhang2019openei,chen2019eyeriss,tan2021efficient,niu2022gcd,prakash2023cfu,yan2023hardware}. 
%Moreover, there are continuous efforts to migrate DL to emerging hardware platforms~\cite{mohammadi2018deep,debauche2020new,vrevca2020accelerating,zhang2019openei,chen2019eyeriss,tan2021efficient,niu2022gcd,prakash2023cfu,yan2023hardware}. 
However, CML and DA have not received as much attention. 
They have limited hardware support and are unable to leverage the advancements in the DL domain.

%To sum up, substantial disparities between software and hardware persist in the realms of deep learning, classical machine learning, and data analysis, constraining the utilization and development of both. 

This paper proposes a systematic methodology for constructing a unified framework that bridges the gap between multiple domain-specific frameworks and diverse hardware devices. The methodology outlined in Fig.~\ref{fig_method} demonstrates how this approach reduces the complexity of porting applications to different hardware platforms, achieving a complexity of O(M+N).
It employs multi-layer abstractions: various domain-specific abstractions are used to represent applications from different domains, which are transformed into a unified abstraction, then lowered to the combinations of primitive operators that are finally mapped to multiple hardware platforms.
It utilizes domain-specific abstractions to support diverse application domains, addresses portability issues through unified abstraction, leverages existing frameworks and compilers to reduce engineering costs, and incorporates multi-level optimizations to guarantee performance.
A unified framework is implemented to support high-level application domains, including DL, CML, and DA, and low-level hardware devices, such as X86, ARM, RISC-V, IoT devices, and GPU. %bridge the gap between high-level applications and low-level hardware devices.
%Hence, this paper proposes a method for constructing machine learning and data analysis frameworks targeting multiple hardware types. 
%It implements a unified framework that supports deep learning, classical machine learning, and data analysis at the upper level while providing support for various hardware architectures such as X86, ARM, and RISC-V at the lower level.

This paper makes the following contributions:

\begin{itemize}
\item We present a systematic methodology for constructing a unified framework that effectively bridges the gap between domain-specific frameworks and hardware devices. This methodology reduces the theoretical complexity of porting applications from  O(M$\times$N) to O(M + N).
\item We design a unified framework that supports three application domains including DL, CML, and DA on various hardware devices such as X86, ARM, RISC-V, IoT devices, and GPU.
\item We implement our work on top of TVM and LLVM, support a broader range of hardware, and achieve a speedup of 1.1x to 3.83x on X86 servers, 1.06x to 4.33x on ARM IoT devices, 1.25x to 3.72x on RISC-V IoT devices, and 1.93x on GPU, compared to existing solutions such as scikit-learn, hummingbird, Spark, and pandas.
\end{itemize}

The remainder of the paper is organized as follows.
Section~\ref{sec_challenge} introduces the challenges.
Section~\ref{sec_method} introduces the method.
Section~\ref{sec_design_implementation} details the design and implementation.
Section~\ref{sec_evaluation} showcases the experimental results.
Section~\ref{sec_related_work} discusses related work.
Finally, we conclude in Section~\ref{sec_conclusion}.

%\item We introduce primitive operator set. 
%Algorithms from the three domains can be represented as the combinations of these primitive operators. 
%A code generator is utilized to implement and optimize these primitive operators, generating high-performance code based on hardware characteristics.

\iffalse
\begin{figure}
    \centering
    \includegraphics[width=0.48\textwidth]{./figures/status.pdf}\\
    \caption{Current domain-specific frameworks lack support and optimizations for diverse hardware devices. Case-by-case implementations have a complexity of O(M$\times$N), increasing the porting cost theoretically.
    }
    \label{fig_status}
\end{figure}
\begin{figure}
    \centering
    \includegraphics[width=0.48\textwidth]{./figures/methodology.pdf}\\
    \caption{We propose a method to systematically bridge the gap between domain-specific frameworks and multiple hardware devices, reducing porting complexity from O(M$\times$N) to O(M+N).
    }
    \label{fig_methodology}
\end{figure}
\fi
\section{Challenges}\label{sec_challenge}
To facilitate the efficient execution of diverse applications from different domains across multiple hardware platforms, three main challenges need to be addressed: portability, performance, and expressiveness.
Section~\ref{sec_portability} introduces portability, Section~\ref{sec_challenge_performance} introduces performance, Section~\ref{sec_expressiveness} introduces expressiveness.

\subsection{Portability}\label{sec_portability}
Portability refers to the ability to transfer a program from one hardware environment to another~\cite{poole1975portability}, and the primary factors limiting portability are hardware differences such as architecture and instruction sets.
%As shown in the figure, there are significant differences in instruction sets among different hardware platforms, resulting in substantial variations in corresponding assembly code. These differences pose significant challenges to portability. 
Efficiently supporting complex applications on diverse and significantly different hardware has always been a hot topic in computer science.

In the early days, developers primarily used assembly language to develop programs on a specific hardware platform~\cite{booth1947coding,wilkes1951preparation}, without considering portability. 
Early compilers~\cite{rutishauser1951automatische,backus1957fortran,10.1145/321439.321440} could translate high-level languages into assembly code, but they were specific to a particular hardware platform and lacked portability. 
With the increasing variety of hardware types, the case-by-case implementation could no longer meet the users' demands, and the portability issue entered the researchers' view.

%In the 1960s, compilers that could support multiple hardware platforms emerged~\cite{rosen1961altac,sammet1978early}, but they were still implemented on a case-by-case basis and not a general solution. 
In the late 1970s, PQCC~\cite{leverett1980overview,wulf1980pqcc,cattell1979code,brosgol1980tcolada} introduced a two-stage structure.
The frontend analyzes syntax and semantics to generate Intermediate Representation (IR), and the backend reads the IR to generate executable code. 
PQCC allowed the same code to run on multiple hardware platforms, ensuring portability.
%, and it was the IR that bridged the gap between the high-level source code and the low-level hardware 
%Subsequent compilers followed similar approaches but supported more programming languages and hardware platforms, and introduced more optimization techniques.
Subsequent compilers such as GCC~\cite{stallman2003using} and LLVM~\cite{lattner2004llvm} introduced a middle-end between the frontend and backend, forming a three-stage structure.
They support more programming languages and hardware platforms and introduce more optimization techniques.
%The optimizer in the middle-end can perform equivalent transformations on the IR, greatly aiding performance improvement.

%GCC (GNU Compiler Collection)~\cite{stallman2003using}, which emerged in the mid-1980s, is one of the most widely used compilers today. 
%It supports a wide range of programming languages, including C, C++, Fortran, and Ada, among others. 
%It also introduced a middle-end between the front-end and back-end, forming a three-stage structure. The optimizer in the middle-end can perform equivalent transformations on the intermediate representation, greatly aiding performance improvement. Furthermore, the introduction of the middle-end further decouples the front-end and back-end. The front-ends for multiple programming languages can generate a common intermediate code, the middle-end performs language- and hardware-independent optimizations, and the back-end performs hardware-specific optimizations to generate high-performance executable code. LLVM~\cite{lattner2004llvm} is another significant advancement in the domain of general-purpose compilers. It follows a similar three-stage structure, but its better modularity allows developers to easily extract specific components and integrate them with other frameworks. Its unified intermediate representation, LLVM Intermediate Representation (LLVM IR), has also been widely adopted.

Looking back at the history of compiler development, we can find that the introduction of the two-stage structure and unified IR marked significant milestones.
When it comes to supporting M programming languages on N hardware platforms, implementing programming language on hardware case-by-case leads to a complexity of O(M $\times$ N). 
%, using the approach of IR significantly reduces complexity and engineering effort. 
By utilizing a unified IR, where various programming languages can be translated into the IR and then used to generate executable code for different hardware platforms, the complexity reduces to O(M + N).
The introduction of the IR bridges the gap between source code and hardware, significantly reducing porting complexity and engineering costs, and finally improving portability.

General-purpose compilers allow users to develop programs and generate executable code for different hardware platforms, effectively addressing the issue of portability.
However, they lack domain-specific information and require users to perform domain-specific optimizations. 
Additionally, users need to implement algorithms from scratch at a low level, which can be quite labor-intensive. 
With the rapid development of emerging application domains such as DL, CML, and DA, users pay more attention to adapting high-level applications and innovating algorithms, hoping to reuse underlying implementations and optimizations. 
General-purpose compilers cannot fully meet these user needs. 
As a result, many domain-specific frameworks have emerged. 
They encapsulate algorithms specific to a particular domain, hiding the low-level implementation details, and making it easier for users to utilize. 
Consequently, these frameworks have gained popularity. 
However, this situation raises the issue of portability once again.

%Domain-specific libraries such as implement and optimize algorithms from scratch, providing them as library functions for users. However, these libraries are limited to specific algorithms and they are primarily optimized for certain hardware, resulting in poor portability. 
%Domain-specific frameworks support a wider range of algorithms. 
%They leverage high-performance libraries, 
%They leverage high-performance libraries that support specific hardware, mainly targeting well-established hardware ecosystems. 
%The portability of these frameworks is constrained by the portability of the underlying high-performance libraries. 
%DL compilers have improved portability by compiling for a broader range of hardware platforms. 
%However, there hasn't been similar progress in the CML and DA domains, which leads to significant portability issues in these two domains.
Porting a framework to a particular hardware typically involves two methods: leveraging high-performance libraries like BLAS~\cite{blackford2002updated} and cuDNN~\cite{chetlur2014cudnn}, or utilizing compilers such as LLVM~\cite{lattner2004llvm}. 
DL frameworks like TensorFlow and PyTorch use high-performance libraries to support hardware like X86, ARM, and GPUs.
%Many specialized hardware in deep learning domains have also aligned with these frameworks by providing libraries.
However, many hardware devices, such as numerous IoT devices, lack the necessary libraries required by these frameworks. 
DL compilers like TVM support such hardware through compilation, by generating LLVM IR and utilizing LLVM to support them. 
%These efforts have largely bridged the gap between software and hardware in the deep learning domain.
However, CML and DL frameworks such as scikit-learn and pandas primarily rely on libraries and are mainly fit for hardware with mature ecosystems like X86 and ARM, while lacking support and optimizations for emerging hardware like RISC-V, GPUs, and IoT devices.
%For classical machine learning and data analysis, the gap between software and hardware is significant.

\subsection{Performance}\label{sec_challenge_performance}

Performance is also an important challenge, especially for latency-sensitive and resource-constrained tasks. 
To ensure performance, multiple levels need to be considered.

At the highest level, the characteristics of the application domain need to be analyzed to fully explore optimization opportunities based on their features. 
Take DL as an example.
%Taking deep learning as an example, many features can be used for optimization. 
Many operators have data dependencies and their mathematical properties ensure that their fusion does not affect the results, leading to operator fusion optimizations~\cite{niu2021dnnfusion,zhao2022apollo,zheng2023chimera}. 
Certain subgraphs in the computational graph can be transformed equivalently, suitable for different scenarios, which gives rise to graph substitutions replacing optimizations~\cite{jia2019taso,ma2020rammer,fang2020optimizing}. 
Data can have different layout formats, such as NCHW, NHWC, and NCHWc, which are suitable for different workloads and hardware, resulting in data layout transformation optimizations~\cite{liu2019optimizing,xing2019dnnvm}. 
%Additionally, there are other optimizations such as graph pruning and tensor recomputation. 
%In summary, the characteristics of the application domain bring significant optimization potential that needs to be fully exploited to improve performance.

At the lowest level, it is crucial to fully leverage hardware features, including processor architecture and instruction sets, parallelism, and memory hierarchy. 
Hardware-related optimizations include loop transformation~\cite{davidson1995aggressive,huang1999generalized,booshehri2013improving,kennedy1993maximizing,fraboulet2001loop,mehta2014revisiting,callahan1988compiling,mahlke1992effective}, parallelization~\cite{subhlok1993exploiting,lilja1994exploiting,reinders2007intel,karlin2013exploring}, vectorization~\cite{eichenberger2004vectorization,nuzman2006multi,nuzman2006auto}, among others.
%Hardware-related optimizations include loop unrolling~\cite{davidson1995aggressive,huang1999generalized,booshehri2013improving}, loop fusion~\cite{kennedy1993maximizing,fraboulet2001loop,mehta2014revisiting}, loop splitting~\cite{callahan1988compiling,mahlke1992effective}, blocking~\cite{panda1999augmenting,renganarayanan2012parameterized,bao2013defensive}, parallelization~\cite{subhlok1993exploiting,lilja1994exploiting,reinders2007intel,karlin2013exploring}, vectorization~\cite{eichenberger2004vectorization,nuzman2006multi,nuzman2006auto}, instruction scheduling~\cite{gibbons1986efficient,wilken2000optimal,faraboschi2001instruction}, and register allocation~\cite{chaitin1981register,poletto1999linear,quintao2008register}, among others.

In addition, many optimization techniques at the intermediate level are unrelated to the application and hardware but are relatively more generic, such as common subexpression elimination~\cite{cocke1970global,potkonjak1996multiple} and dead code elimination~\cite{knoop1994partial,xi1999dead}.
%These include constant propagation~\cite{callahan1986interprocedural,wegman1991constant}, common subexpression elimination~\cite{cocke1970global,potkonjak1996multiple}, dead code elimination~\cite{knoop1994partial,xi1999dead}, and inlining~\cite{dean1994towards,ayers1997aggressive}, among others. 
To ensure performance, optimization methods at all these levels should not be overlooked.

%Many optimizations in domain-specific frameworks for DL, CML, and DA rely on high-performance libraries and general-purpose compilers. 
High-performance libraries encapsulate common computations and provide user-friendly programming interfaces after manual tuning. 
Many hardware vendors also provide high-performance libraries targeting their hardware~\cite{wang2014intel,armcomputelib,cublas,cutensor,chetlur2014cudnn}.
These high-performance libraries cover low-level and intermediate-level optimizations but do not involve domain-specific optimizations.
Currently, many domain-specific frameworks call high-performance libraries and supplement them with domain-specific optimizations. 
The problem with high-performance libraries is that, on the one hand, they do not support comprehensive computations and often cannot meet the requirements of high-level application domains. 
On the other hand, they have limited hardware support, lacking optimizations for many emerging hardware platforms.

General-purpose compilers like GCC and LLVM cover low-level and intermediate-level optimizations but do not involve domain-specific optimizations. 
They require high-level frameworks to provide domain-specific optimizations. 
Currently, these three application domains do not directly utilize general-purpose compilers to support high-level applications but rather use them as backends to efficiently support diverse hardware platforms. 
However, the automatic optimizations of general-purpose compilers are not comprehensive, and they require optimization-related information from high-level frameworks.

DL frameworks like TensorFlow and PyTorch have unified intermediate abstractions that allow the reuse of many domain-specific optimizations. 
This enables them to fully leverage domain-specific features.
The intermediate and low-level optimizations of these frameworks rely on high-performance libraries, facing challenges in optimizing for emerging hardware. 
DL compilers like XLA and TVM follow the same high-level optimizations but interface with general-purpose compilers like LLVM at the lower level. 
They utilize these compilers for intermediate and low-level optimizations, supporting a wider range of hardware platforms compared to high-performance libraries and achieving better performance on some hardware.

CML frameworks like scikit-learn and Spark MLlib lack intermediate abstractions, and domain-specific optimizations are performed during algorithm implementations.
This makes it difficult to reuse optimizations for different algorithms, which impacts performance and increases engineering costs. 
Intermediate and low-level optimizations rely entirely on high-performance libraries but are also limited by them. 

DA frameworks like Spark and pandas have unified intermediate abstractions that allow the reuse of domain-specific optimizations to fully leverage domain-specific features. 
They also rely on high-performance libraries and are therefore constrained by them.
In summary, ensuring performance across multiple hardware platforms is an important challenge, related to multi-level optimizations.

\subsection{Expressiveness}\label{sec_expressiveness}
To support high-level applications, the framework needs to have strong expressiveness, which involves clear and accurate data abstraction, computation abstraction, and representation of data flow and computation dependencies.
%Expressiveness mainly encompasses three aspects: data abstraction, computation abstraction, and representation of data flow and computation dependencies.
%To support high-level applications, the framework needs to have strong expressiveness, which involves clear and accurate representation. 
%Expressiveness mainly encompasses three aspects: data abstraction, computation abstraction, and representation of data flow and computation dependencies.

Data abstraction encompasses the comprehensive representation of the data used in the applications and the effective organization and manipulation of this data.
Computation abstraction requires an accurate description of the computations within the applications, breaking down complex applications into simpler and more manageable computational units that can be combined.
Data flow and computation dependencies are commonly represented using graphs, with the most common ones being data flow graphs~\cite{1676696} and control flow graphs~\cite{10.1145/390013.808479}.

Due to the variations in data and computations across different application domains, their data and computation abstractions differ as well. %in different domains. 
We discuss DL, CML, and DA separately.
The data used in DL includes images, text, audio, and more. 
Most of these data types are represented, stored, and processed using tensors, which are the fundamental data abstraction for DL. 
In DL, neural network models usually comprise multiple layers, each performing various computational operations such as convolution, matrix multiplication, and activation functions. These operations, known as operators, serve as the fundamental computation abstraction. High-level applications are built by combining these operators in different ways. Computational graphs, represented as directed acyclic graphs (DAGs), are used to represent the data flow and computation dependencies. In these graphs, nodes represent operators, and edges represent tensors.

The data used in CML includes tables, text, images, and more. 
The data abstraction in CML is relatively complex and includes scalars, matrices, arrays, and tables. 
Applications in CML are composed of multiple models organized in pipelines, including data preprocessing, linear models, trees, forests, and more. 
These models are relatively complex but are still considered as the minimum units for implementation and scheduling. 
Pipelines are represented using DAGs, where nodes represent models and edges represent different types of data.

DA primarily deals with data such as tables, text and time series. 
The data abstraction in DA includes relational tables, scalars, and arrays. %Resilient Distributed Datasets (RDDs), and DataFrames. 
Applications in DA are composed of pipelines that combine various operators, including relational operators like selection, projection, and join, as well as basic mathematical and statistical operators like matrix inverse and covariance. 
These operators serve as the computation abstraction. 
Pipelines in DA are represented using DAGs, where nodes represent operators and edges represent data.

We find that the data flow and computation dependencies in these application domains can all be represented using DAGs. 
However, the data and computation abstractions differ significantly. 
The challenge lies in how to represent applications from these different domains and how to represent cross-domain applications effectively.

\section{Methodology}\label{sec_method}
This section introduces the methodology to systematically bridge the gap between domain-specific frameworks and multiple hardware devices, as shown in Fig.~\ref{fig_method}.
We use domain-specific abstractions and unified abstractions to bridge applications and hardware, reducing porting complexity from O(M$\times$N) to O(M+N).

\subsection{Address Portability Issues through Intermediate Abstraction}

Looking back at the development history of general-purpose compilers and the ecosystems of the three application domains, it can be observed that the primary method for addressing portability is through the introduction of intermediate abstractions, bridging the gap between high-level application domains and low-level hardware, thus reducing the porting costs from O(M $\times$ N) to O(M + N).
Based on a similar concept, we propose a unified intermediate abstraction for the three application domains, which can represent high-level algorithms and is then mapped to various low-level hardware platforms.
%As shown in Figure~\ref{fig_methodology_how-to-solve-portabiltiy}, based on a similar concept, this paper proposes a unified intermediate layer abstraction for the three application domains, where algorithms specific to each domain can be represented at the higher level and mapped to various hardware platforms at the lower level.

%Furthermore, because the deep learning ecosystem has better portability among the three application domains, this work leverages the support of deep learning frameworks and compilers for hardware. However, deep learning frameworks also lack support for certain hardware platforms like RISC-V. Therefore, this work complements the lower-level code generator to generate LLVM IR specifically for hardware platforms such as RISC-V, utilizing LLVM's support for these hardware platforms.

\subsection{Propose Domain-specific Abstractions and Unified Abstraction for High-level Applications}

\begin{table*}
	\caption{The differences in Classical Machine Learning, Deep Learning and Data Analysis}   \label{table_domain_difference}
	\center
	{
		\begin{tabular}
			{m{2.4cm}<{\centering}|m{4.5cm}<{\centering}|m{4.5cm}<{\centering}|m{4.5cm}<{\centering}}
			\hline
			& Deep Learning & Classical Machine Learning & Data Analysis \\ \hline
			Data Structure & tensor & scalar, array, table &  relational table, scalar, array \\ \hline
			Data Type & numeric & numeric & numeric, non-numeric \\ \hline
			Computation Abstraction & Operators as the smallest unit; Algorithms represented as the combinations of operators & No computation abstraction; Case-by-case implementation; Large number of models & Operators as the smallest unit; Small number of operators  \\ \hline
			%领域特性 & & 天然具有稀疏性；流水线中数据的精度会发生变化 & 许多算子满足交换律或结合律\\ \hline
	\end{tabular}}
\end{table*}

The intermediate abstractions need to be able to represent applications from different domains. 
We propose a two-layer abstraction: domain-specific abstractions for interfacing with application domains, which are then transformed into a unified abstraction for interfacing with hardware. 
The unified abstraction can also be used to represent cross-domain applications. 

%The essence of intermediate layer abstraction is to extract commonalities from higher-level applications. The more commonalities there are, the lower the porting costs.
First, we introduce the unified abstraction, which requires the exploration of commonalities in three application domains.
The biggest commonality among DL, CML, and DA is that they can all be represented as DAGs to depict data flow.
This forms the basis for cross-domain pipelines and forms the establishment of a unified abstraction. 
Therefore, we propose a unified abstraction based on DAG.

%Additionally, all three domains heavily use calculations such as matrix multiplication, mean, variance, etc., which can be reused in their implementations and optimizations.

The meanings of nodes and edges in the corresponding DAGs of the three domains differ due to the differences in data and computation abstractions. 
To accurately represent applications from the three domains, we propose domain-specific abstractions for each domain. 
As shown in Table~\ref{table_domain_difference}, for DL, data mainly consists of numerical tensors and the computation abstraction treats DL operators as the smallest unit, with algorithms represented as the combinations of operators. 
We adopt the existing DL abstractions, using tensors as the data abstraction and DL operators as the computation abstraction.

For CML, data mainly consists of numerical scalars, arrays, and tables, all of which can be converted to tensors directly.
However, there is no suitable computation abstraction. 
Models are implemented case by case, and there are a large number of models. 
The problem lies in computation abstraction.
We use classical operator representation (COR)~\cite{wen2023cmlcompiler} as the computation abstraction of CML.
It represents CML models as combinations of assignment, exchange, basic arithmetic, aggregation, comparison, and conditional operators.
In summary, we use tensors as the data abstraction and COR as the computation abstraction for CML.

For DA, data structures include relational tables, scalars, and arrays, with both numerical and non-numerical data types. 
The computation abstraction in DA also treats operators as the smallest unit, and there are only a small number of operators. 
The problem lies in data abstraction.
We use TensorTable~\cite{wen2024tensortable} as the data abstraction of DA, it stores data in a tensor format and can represent different data structures and data types mentioned above.
In summary, we use TensorTable as the data abstraction and DA operators as the computation abstraction for DA.

With the above approach, appropriate domain-specific abstractions are constructed for different application domains.
Subsequently, three domain-specific abstractions are transformed into the unified abstraction. 
The data abstraction of the unified abstraction includes tensor and TensorTable, while the computation abstraction includes DL operators, COR, and DA operators.

\subsection{Reuse Existing Frameworks and Compilers to Reduce Engineering Cost}
We use domain-specific abstractions and unified abstractions to represent high-level applications.
However, to run multiple applications on different hardware devices, these abstractions need to be mapped to the hardware.
We leverage existing frameworks and compilers to support multiple hardware platforms, thereby reducing the engineering cost. 
Due to the better portability of DL compilers compared with other domain-specific frameworks of three domains, we reuse DL compilers to support hardware such as X86, ARM, GPUs, and IoT devices.
However, DL compilers also lack support for certain hardware platforms like RISC-V. 
Therefore, we introduce a low-level code generator to generate LLVM IR specifically for RISC-V and similar hardware, leveraging LLVM's support for these hardware.
Additionally, DL compilers rely on auto-tuning for optimizations, which incurs high time costs and are not suitable for CML and DA. 
Therefore, we propose new optimization methods specifically for them.

\subsection{Use Multi-level Optimizations to Guarantee Performance}
To improve performance, it is necessary to fully utilize the characteristics of application domains and leverage hardware capabilities. 
Our work mainly includes two-level optimizations: graph-level optimizations and operator-level optimizations.

%As shown in Table~\ref{table_graph_optimization}, g
Graph-level optimization can be categorized into three types. 
Domain-specific optimizations are tailored to the characteristics of each application domain and are unique. 
To capture the domain-specific features, we propose domain-specific abstractions for the three application domains and conduct domain-specific optimization based on these abstractions, such as operator fusion and data layout transformation for DL, sparse operator replacing and data type rewriting for CML, as well as operator reorder and operator fusion for DA.
%Domain-specific optimizations for deep learning include operator fusion, subgraph replacement, data layout transformation, graph pruning, and tensor recomputation. Domain-specific optimizations for traditional machine learning include sparse operator replacement, data type rewriting, and redundant operator elimination. Domain-specific optimizations for data analysis include operator reordering and operator fusion.
After domain-specific optimizations, three domain-specific abstractions are transformed into a unified abstraction. 
The unified abstraction enables cross-domain optimizations for mixed workloads from three domains, primarily including cross-domain operator reorder and cross-domain operator fusion.
Subsequently, generic graph-level optimizations are applied to the unified abstraction, such as constant propagation and common subexpression elimination.

After graph-level optimizations are completed, operator-level optimizations are performed.
%As shown in Table~\ref{table_operator_optimization}, 
Operator-level optimization mainly consists of two types: hardware-agnostic optimizations such as constant folding and copy propagation, and hardware-specific optimizations such as loop transformation and vectorization.
All those optimizations work together to guarantee performance.

\section{The System Design and Implementation}\label{sec_design_implementation}
This section illustrates the design and implementation.
Section~\ref{sec_design} shows the design overview.
Section~\ref{sec_domain_abstraction} introduces domain-specific abstractions and optimizations.
Section~\ref{sec_unified_abstraction} introduces unified abstraction and optimizations.
Section~\ref{sec_primitive_operator} introduces primitive operator set.
Section~\ref{sec_codegen} introduces code generation and operator-level optimizations.
Section~\ref{sec_arithecture} shows the implementation.

\subsection{Design Overview} \label{sec_design}
\begin{figure}
    \centering
    \includegraphics[width=0.45\textwidth]{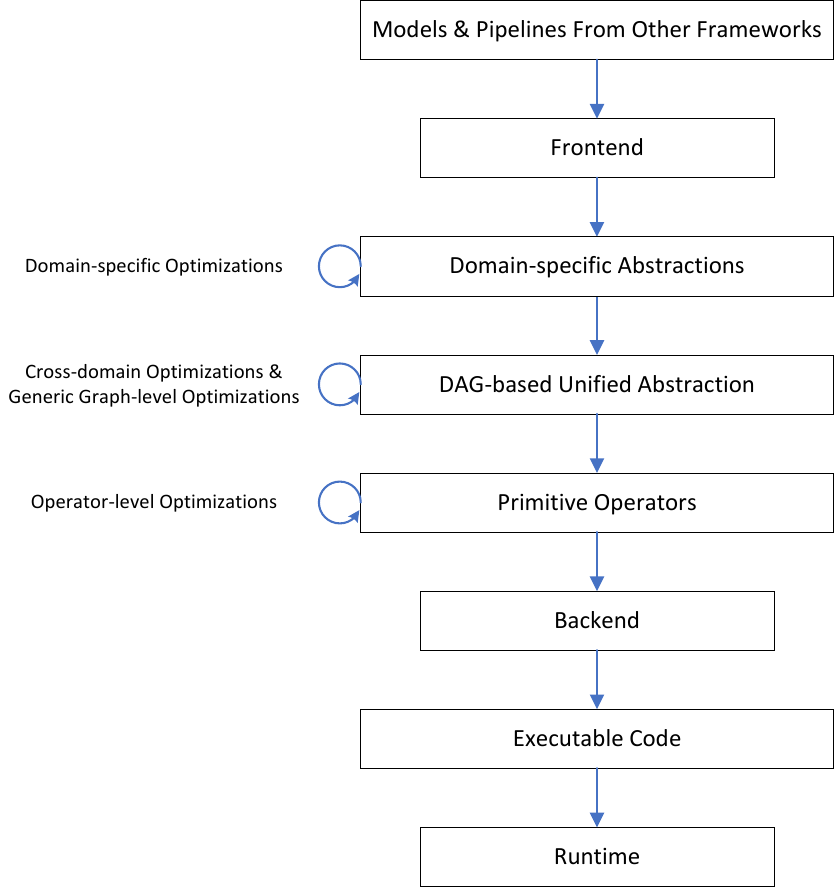}
    \caption{Design Overview.
    }
    \label{fig_design}
\end{figure}
Fig.~\ref{fig_design} shows the design overview.
The frontend is used to read models and pipelines from other frameworks and convert them to domain-specific abstractions.
Cross-domain applications that use multiple frameworks are converted to many subgraphs and stitched together based on data dependencies.
Domain-specific abstractions are used to represent different application domains, capturing their data and computation features.
They are converted to a DAG-based unified abstraction, which is then mapped to the combination of primitive operators.
Multi-level optimizations are performed for those abstractions, including domain-specific optimizations for domain-specific abstractions, cross-domain and generic graph-level optimizations for DAG-based unified abstraction, as well as operator-level optimizations for primitive operators.
The backend is used to generate executable code.
The runtime reads executable code and input data, running on multiple hardware.

\subsection{Domain-Specific Abstractions and Optimizations}\label{sec_domain_abstraction}
We use domain-specific abstractions to represent high-level applications, all of which are based on DAG, with different data and computation abstraction.
For DL, we use tensors as the data abstraction and DL operators as the computation abstraction.
For CML, we use tensors as the data abstraction and classical operator representation (COR)~\cite{wen2023cmlcompiler} as the computation abstraction.
COR uses the combinations of six categories of operators, namely assignment, swap, basic arithmetic, aggregation, comparison, and conditional operators, to represent widely-used CML algorithms, such as preprocessing algorithms, linear models, trees and forests, and SVMs.
For DA, we use TensorTable~\cite{wen2024tensortable} as the data abstraction and DA operators as the computation abstraction.
TensorTable represents relational tables as a combination of tensors and dictionaries, by converting those non-numerical data into numerical data and using auxiliary dictionaries to preserve the mapping relations.
TensorTable can also seamlessly encapsulate other data structures such as scalars and arrays.
DA operators are all implemented based on TensorTable, including relational operators, covering selection, projection, join, group by, and aggregation, as well as basic mathematical and statistical operations.

%Domain-specific optimization is tailored to the characteristics of each application domain and is unique to each domain. To capture the domain-specific features, this work proposes domain-specific abstractions for the three application domains and conducts domain-specific optimization based on these abstractions.
We conduct domain-specific optimizations based on these abstractions to fully utilize the features of each domain, as shown in Table~\ref{table_domain_optimization}.
Domain-specific optimizations for DL include operator fusion, data layout transformation, graph substitutions replacing, and tensor recomputation.
Domain-specific optimizations for CML include sparse operator replacing, data type rewriting, and redundant operator elimination. 
Domain-specific optimizations for DA include operator reorder and operator fusion.

\begin{table*}
	\caption{Domain-specific optimizations for domain-specific abstractions}   \label{table_domain_optimization}
	\center
	{
		\begin{tabular}
			{m{6cm}<{\centering}|m{6cm}<{\centering}|m{4cm}<{\centering}}
			\hline
			\textbf{Deep Learning}& \textbf{Classical Machine Learning} & \textbf{Data Analysis} \\ \hline
			Operator fusion, Data layout transformation, Graph substitutions replacing, Tensor recomputation & Sparse operator replacing, Data type rewriting, Redundant operator elimination & Operator reorder, Operator fusion \\ \hline
	\end{tabular}}
\end{table*}

\subsection{Unified Abstractions and Optimizations}\label{sec_unified_abstraction}
%Domain-specific abstractions serve to interface with higher-level applications, preserving the unique characteristics of each application domain while optimizing related aspects.

%For deep learning, the widely used computational graph remains the domain-specific abstraction. It functions as a directed acyclic graph (DAG), with nodes representing deep learning operators and edges denoting tensor-formatted data.

%Traditional machine learning utilizes the abstraction introduced in as its domain-specific abstraction. It's also a DAG, with nodes representing traditional operators and edges denoting tensor-formatted data.

%Data analysis adopts the abstraction introduced in as its domain-specific abstraction. Similar to the previous two, it's a DAG, with nodes representing data analysis operators and edges denoting tensor tables.

%It's notable that all three domain-specific abstractions can be represented as DAGs, differing only in the meaning of nodes and edges. Hence, a directed acyclic graph can be used as a unified abstraction. It includes tensors required for both deep learning and traditional machine learning, along with tensor tables required for data analysis. The supported operators encompass the aforementioned three domains. This unified abstraction, after optimization, is mapped to multiple hardware platforms, achieving the goal of supporting various hardware while ensuring performance.

After domain-specific optimizations, all three domain-specific abstractions are transformed into a unified abstraction. 
It's also based on DAG, with nodes representing operators and edges denoting data.
Its data abstraction includes both tensor and TensorTable, and its computation abstraction includes DL operators, COR, and DA operators.

As shown in Table~\ref{table_graph_optimization}, unified abstraction enables many optimizations, including cross-domain optimizations and generic graph-level optimizations.
Cross-domain optimizations cater to mixed workloads from three domains, primarily including cross-domain operator reoder and cross-domain operator fusion.
Cross-domain operator reorder changes the order of operators from different domains which meet commutativity, to decrease computation.
Cross-domain operator fusion merges operators from different domains into one operator, to reduce intermediate data. 
Generic graph-level optimizations include constant folding, constant propagation, copy propagation, common subexpression elimination, dead code elimination, and more.
\begin{table}
	\caption{Graph-level optimizations for unified abstraction}   \label{table_graph_optimization}
	\center
	{
		\begin{tabular}
			{m{8cm}<{\centering}} \hline
			\textbf{Cross-domain Optimizations} \\ \hline
			Cross-domain operator reorder, Cross-domain operator fusion \\ \hline
                \textbf{Generic Optimizations} \\ \hline
                Constant folding, Constant propagation, Copy propagation, Common subexpression elimination, Dead code elimination, etc.\\ \hline
	\end{tabular}}
\end{table}

\subsection{Primitive Operator Set}\label{sec_primitive_operator}
After graph-level optimizations, the unified abstraction is mapped down to the combinations of operators.
DL encompasses over two thousand operators\footnote{\url{https://pytorch.org/get-started/pytorch-2.0/\#primtorch-stable-primitive-operators}}, while CML and DA each involve dozens of operators.
The cost of implementing and optimizing operators one by one is prohibitively high. 
Therefore, we construct a primitive operator set consisting of eighty operators, focusing on their implementations and optimizations. 
Algorithms used by high-level applications can be represented as the combinations of them.
Subsequently, the primitive operator set is implemented, generating executable code tailored for multiple hardware platforms.

The process of constructing the primitive operator set is as follows:

(1) We summarize all required operators in DL, CML, and DA, totaling around 2300 operators.
(2) We eliminate long-tail operators that are rarely used and those various overloads for the same operators, reducing to around 300 operators.
(3) We retain operators that cannot be formed by combinations of other operators or those whose performance is significantly lower when combined, resulting in 80 operators, as the primitive operator set.
The final primitive operator set is shown in Table~\ref{table_primitive}.

\begin{table}
    \caption{Primitive operators}
    \label{table_primitive}
    \center
    {
        \begin{tabular}
	{|p{8cm}<{\centering}|}
	\hline
        \textbf{Injective operators}\\ \hline
  	mod, floor, ceil, clip, round, abs, negative, pow, log, sqrt, exp, sin, cos, tan, not \\ \hline
        \textbf{Element-wise operators}\\ \hline
        add, subtract, multiply, divide, equal, not\_equal, greater, greater\_equal, less,   
        less\_equal, all, any, and, or, xor, where \\ \hline
        \textbf{Reduction operators}\\ \hline
        max, min, sum, mean, topk, cumsum, cumprod \\ \hline
        \textbf{Index operators} \\ \hline
        nonzero, argmax, argmin, argsort, argwhere \\ \hline
        \textbf{Memory operators}\\ \hline
        reshape, copy, transpose, squeeze, take, cast, tile, reverse, slice, scatter, split, 
        flatten, gather, expand, concatenate \\ \hline 
        \textbf{Key construct opertors} \\ \hline
        inner\_key\_construct, outer\_key\_construct, left\_key\_construct,  
        right\_key\_construct, cross\_key\_construct \\ \hline
        \textbf{Special computation operators}\\ \hline
        matmul, conv, max\_pool, avg\_pool, one\_hot, resize, pad, dropout, unique, sort, 
        intersection, union, complement \\ \hline
        \textbf{Control operators}\\ \hline
        if\_then\_else, loop, while, scan\\ \hline        
    \end{tabular}}
\end{table}

Primitive operators are primarily divided into eight categories:
Injective operators perform identical computations on each element within a single tensor, including operations like modulus (mod), negation (negative), and absolute value (abs), among others.
Element-wise operators perform operations between corresponding elements in two tensors, such as addition (add), subtraction (subtract), multiplication (multiply), division (divide), and so on.
Reduction operators derive aggregated results from a single tensor across one or multiple dimensions. 
Examples include maximum value (max), minimum value (min), summation (sum), and mean computation (mean), among others.
Index operators compute each element within a single tensor and return the corresponding index. 
Examples include returning indices for non-zero values (nonzero), returning indices for maximum values (argmax), and indices for minimum values (argmin), among others.
Memory operators manage memory-related actions, such as reshaping (reshape), copying data (copy), transposing (transpose), and others.
Key construct operators are used in join operations to build key-value indices, including inner\_key\_construct to construct keys for inner join, outer\_key\_construct to construct keys for outer join, and others.
Special computation operators encompass complex and crucial computations like matrix multiplication (matmul) and convolution (conv).
While these operations can be expressed as the combinations of injective, element-wise, and reduction operators, they are retained separately to ensure optimizations, especially for hardware-specific optimizations.
Control operators represent control flow within computations, such as if condition (if\_then\_else) and loop.

\subsection{Code Generation and Operator-level Optimizations}\label{sec_codegen}

For primitive operators, we adopt scheduling techniques proposed by Halide~\cite{ragan2013halide} and TVM to describe and optimize them.
The utilized schedules are outlined in Table~\ref{table_schedule}.
Optimizations such as parallelization, memory hierarchy, and vectorization are employed based on hardware specifications to ensure performance. 
%Subsequently, LLVM IR is generated, utilizing LLVM to generate executable code across multiple hardware platforms.
For hardware such as X86, ARM, and GPUs that are well supported by TVM, the code generation part of TVM is directly reused to generate executable code.
To support more hardware platforms such as RISC-V, we supplement code generation modules specifically for them, which can generate LLVM IR that can be further optimized and translated into executable code using LLVM.
Parallelization is directly implemented using relevant APIs like llvm::parallel. 
Vectorization is achieved through hardware-specific LLVM intrinsics, for instance, $llvm.riscv.v$ for RISC-V vector extension instructions.

\begin{table}
    \caption{Schedules used in Primitive Operators}
    \centering
    \label{table_schedule}
    \center
    {
        \begin{tabular}
	{|m{3cm}<{\centering}|m{4.5cm}<{\centering}|}
	\hline
        Schedule & Explanation \\ \hline 
        split & Split one axis into multiple axes\\ \hline
        tile& Execute the computation tile by tile over two axes\\ \hline
        fuse& Merge multiple axes into one\\ \hline
        reorder & Change the orders of axes\\ \hline
        compute\_at & Compute at certain axis\\ \hline
        parallel& Use multi-threading\\ \hline
        vectorize/tensorize& Use vectorization or tensorization instructions\\ \hline
    \end{tabular}}
\end{table}

Regarding cache-related optimizations, since many hardware cannot directly manage various levels of cache, optimization is achieved by adjusting parameters within the schedule, such as tile size and split size.
We make cache-related optimizations by solving a constrained optimization problem:

\begin{equation}
\label{equation_1}
\begin{aligned}
\mathop{argmin}\limits_{conf}(\mathop{max}\limits_{l\in L}(\frac{DataMove_l}{BandWidth_l})) \\
\ .s.t. \ \forall l \in L \ DataOccupy_l < Capacity_l\\
\end{aligned}
\end{equation}

The constraints ensure that the data occupation of each cache level is smaller than its corresponding cache capacity. 
The optimization objective is to minimize data movement time.
The symbols are listed in Table~\ref{table_symbol}.

\begin{table}
    \caption{Symbols in Equation ~\ref{equation_1}.}
    \label{table_symbol}
    \center
    {
        \begin{tabular}
	{|m{2cm}<{\centering}|m{6cm}<{\centering}|}
	\hline
        Symbol & Explanation \\ \hline
        $conf$ & The configurations in schedules \\ \hline
        $DataMove_l$ & The data movement of $l$-level cache \\ \hline
        $DataOccupy_l$ & The data occupation of $l$-level cache \\ \hline
        $L$ & Cache level \\ \hline
        $BandWidth_l$ & The bandwidth of $l$-level cache \\ \hline
        $Capacity_l$ & The capacity of $l$-level cache \\ \hline
    \end{tabular}}
\end{table}

\begin{table}
	\caption{Operator-level Optimizations for Primitive Operators}   \label{table_operator_optimization}
	\center
	{
		\begin{tabular}
			{m{8cm}<{\centering}} \hline
			\textbf{Hardware-agnostic Optimizations} \\ \hline
		      Inline, Constant folding,  Constant propagation, Copy propagation, Common subexpression elimination, Dead code elimination, etc.\\ \hline
                \textbf{Hardware-specific Optimizations}\\ \hline
                Loop unrolling, Loop fusion, Loop splitting, Blocking, Data alignment, Parallelization, Vectorization, Register allocation, Instruction scheduling, etc.\\ \hline
	\end{tabular}}
\end{table}

During code generation, we make operator-level optimizations, mainly consisting of two types: hardware-agnostic optimizations and hardware-specific optimizations, as shown in Table~\ref{table_operator_optimization}.
Hardware-agnostic optimization includes inlining, constant folding, constant propagation, copy propagation, common subexpression elimination, dead code elimination, and more.
Hardware-specific optimization includes loop unrolling, loop fusion, loop splitting, blocking, data alignment, parallelization, vectorization, tensorization, register allocation, instruction scheduling, and more.
Many graph-level optimizations and operator-level optimizations share similar names and approaches, but they are targeted at different entities (graphs and operators) and have significant differences in implementations.

Finally, we deploy executable code on lightweight runtime to run DL, CML, and DA applications, as well as cross-domain mixed applications, on multiple hardware platforms.

\subsection{Implementation}\label{sec_arithecture}

\begin{figure}
    \centering
    \includegraphics[width=0.48\textwidth]{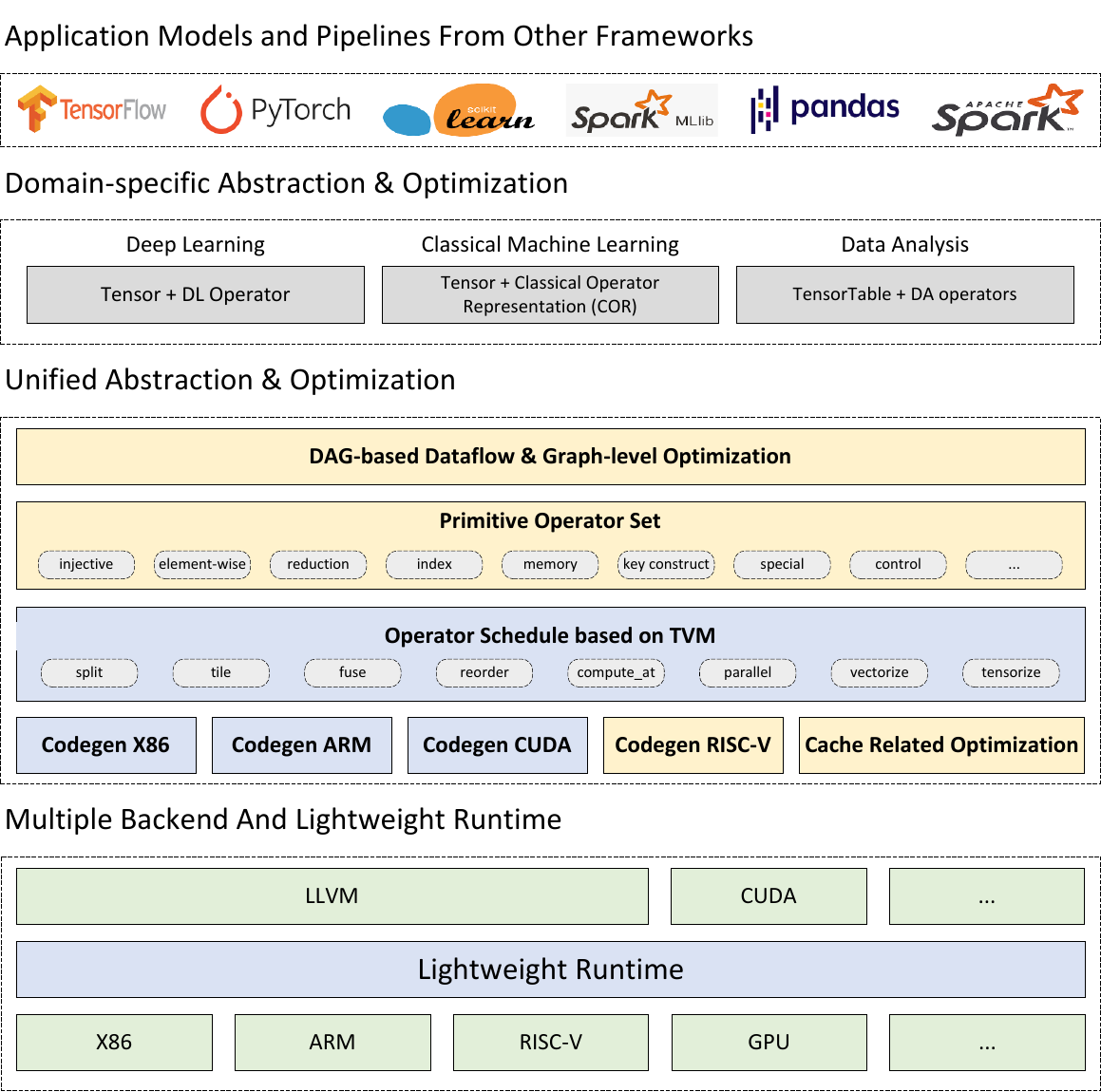}
    \caption{System Architecture.
    }
    \label{fig_framework}
\end{figure}

The system architecture is shown in Fig.~\ref{fig_framework}.
The blue part represents the components reused from TVM, the green part represents the components reused from LLVM and CUDA, and the yellow part represents our contribution.
At the top level, there are application models and pipelines from other frameworks, which are transformed into three domain-specific abstractions using APIs like $from\_pytorch$ and $from\_sklearn$.
After domain-specific optimizations, the three domain-specific abstractions are transformed into a DAG-based unified abstraction. 
The unified abstraction, after graph-level optimization, is mapped down to the combinations of primitive operators.
The primitive operators are implemented using TVM, adopting the decoupled compute/schedule principle~\cite{ragan2013halide,10.5555/3291168.3291211}, striving to reuse code as much as possible while ensuring comprehensive optimization.
Hardware-agnostic and hardware-specific optimizations are performed at the operator level.
Subsequently, executable code is generated for multiple backends.
For hardware such as X86, ARM, and GPUs that are well supported by TVM, the code generation part of TVM is directly reused to generate executable code.
To support more hardware platforms such as RISC-V, we supplement code generation modules specifically for them, which can generate LLVM IR that can be further optimized and translated into executable code using LLVM.
We make cache-related optimizations by solving a constrained optimization problem instead of auto-tuning used by TVM, better fitting for CML and DA workloads.
Finally, the generated code runs on lightweight runtime which is implemented based on TVM runtime and executes on multiple hardware platforms.

\section{Evaluation}\label{sec_evaluation}
This section summarizes the evaluation. 
Section~\ref{sec_setup} shows the experimental setup.
Section~\ref{sec_coverage} evaluates the coverage of primitive operators.
Section~\ref{sec_performance} evaluates the performance on diverse hardware.

\subsection{Experimental Setup}\label{sec_setup}

%This section evaluates the performance of machine learning and data analysis on 
We evaluate the performance on X86 server, ARM IoT device, RISC-V IoT device, and GPU.
We deploy a server node equipped with two Xeon E5-2620 V3 (Haswell) CPUs, an Nvidia Titan RTX GPU, and 64 GB memory to
conduct the experiments on x86 server and GPU. 
Each CPU contains 6 physical cores with hyper-threading. 
The GPU contains 4608 CUDA cores and 24 GB memory, CUDA version is 10.2.
The operating system used is Ubuntu 16.04.
The ARM IoT device used is the Raspberry Pi 4B, which features a quad-core Cortex-A72 CPU and 4GB memory. 
It runs on the Raspbian 10 operating system.
The RISC-V IoT device employed is the VisionFive 2, equipped with a quad-core JH7110 CPU and 8GB memory. 
It runs on the Debian 12 operating system.
The software utilized includes TVM 0.15, scikit-learn 1.1.2, Intel(R) Extension for Scikit-learn 2021.20220215.162512, hummingbird 0.3.1, CMLCompiler 1.01, pandas 1.3.5, Spark 3.3.0, and MonetDB 11.39.17.

\subsection{The Coverage of Primitive Operators}\label{sec_coverage}
\begin{figure}
    \centering
    \includegraphics[width=0.49\textwidth]{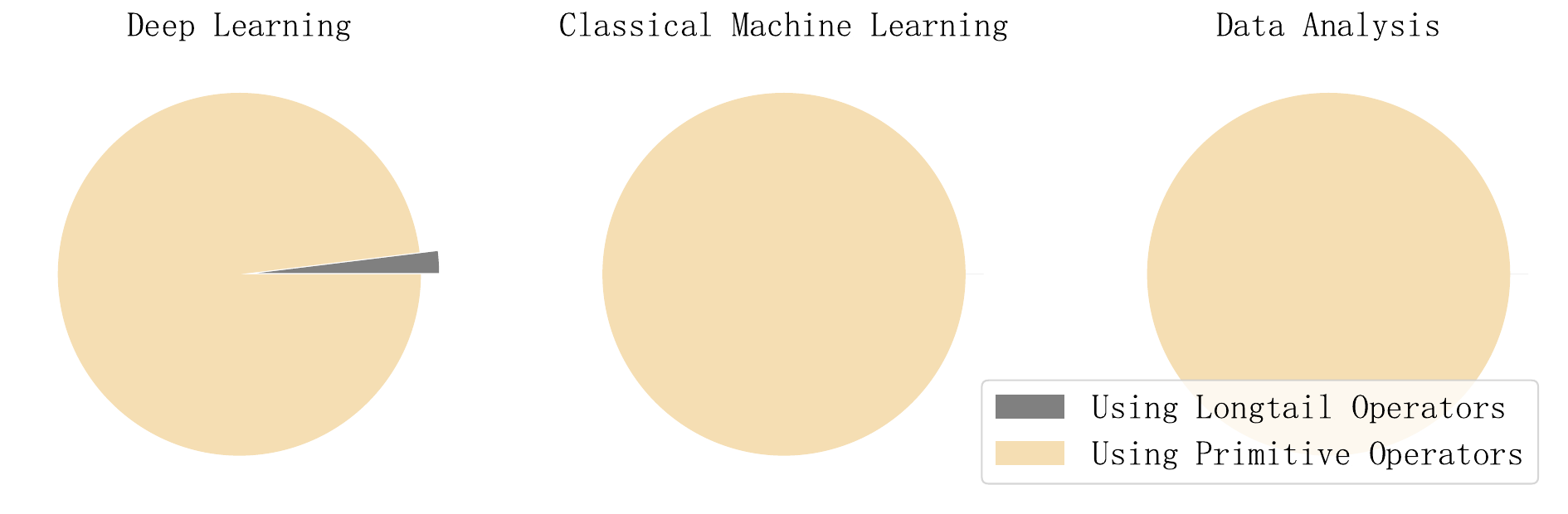}
    \caption{The coverage of primitive operators.
    }
    \label{fig_op_coverage}
\end{figure}

First, we analyze the coverage of the primitive operator set.
To begin, we analyze the coverage within the DL domain. 
A set of 103 widely-used DL models is extracted from the ONNX model repository\footnote{\url{https://github.com/onnx/models}}.
Among them, 101 models were entirely composed of primitive operators, while only two models contained long-tail operators. 
Specifically, the ResNet-preproc model has the $sequenc\_map$ operator, and BiDAF model has the $category\_mapper$, $hardmax$, and $compress$ operators, which are long-tail operators.
Next, we analyze the coverage within the CML domain. 
We extract 47 high-frequency models shared between two widely used frameworks scikit-learn and Spark MLlib. 
All these models can be entirely represented by primitive operators, without any occurrences of long-tail operators.
Finally, we analyze the coverage within the DA domain. 
We analyze 22 DA pipelines from TPC-H\footnote{\url{https://www.tpc.org/tpch/}}.
All of them can be fully represented by primitive operators, without any long-tail operator occurrences.
In summary, as illustrated in Fig.~\ref{fig_op_coverage}, 98\% of the 103 DL models, all 47 CML models, and all 22 DA pipelines can be represented by the primitive operators. 
This demonstrates that the primitive operator set adequately covers the majority of scenarios in DL, CML, and DA applications.

\subsection{Performance} \label{sec_performance}

\begin{figure*}
    \centering
    \includegraphics[width=0.98\textwidth]{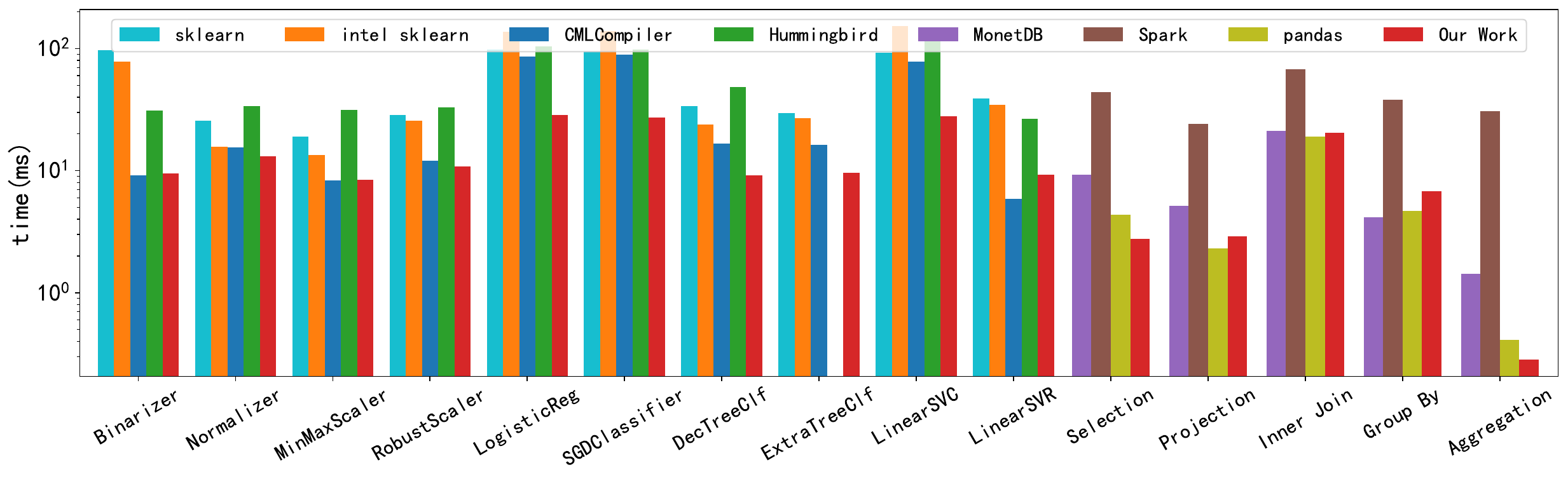}
    \caption{The performance on X86 server. The absent data means unsupported.
    }
    \label{fig_x86}
\end{figure*}

This section compares the performance on various hardware platforms against existing frameworks.
We test ten CML models and compare our work with sklearn, intel sklearn, CMLCompiler, and hummingbird.
Use YearPrediction dataset~\cite{Dua:2019}, which consists of 51,630 samples and 90 features. 
We test five DA operators and compare our work with MonetDB, Spark, and pandas.
Use BIXI dataset~\cite{faghih2014land} at data scales of 1k, 10k, 100k, 1m, and 10m rows, and show their geometric means.
All experiments are repeated five times, and the arithmetic mean is used as the final result.

Fig.~\ref{fig_x86} displays the performance on x86 server.
Our work achieves the best performance in 7 out of 10 tested CML models.
CMLCompiler attains the best performance in 3 models, while sklearn, intel sklearn, and Hummingbird do not outperform in any algorithm. 
We achieve a 1.95x to 10.21x speedup compared with sklearn across 10 models, with an average speedup of 3.83. 
We achieve a 1.2x to 8.18x speedup compared with intel sklearn across 10 models, with an average speedup of 3.79. 
We achieve a 2.57x to 5.25x speedup compared with Hummingbird across 10 models, with an average speedup of 3.24. 
This is attributed to our better utilization of multi-core parallelism and SIMD instructions.
We achieve a 1.11x to 3.25x speedup compared with CMLCompiler across 7 models, with an average speedup of 1.75. 
For algorithms like Binarizer, Normalizer, MinMaxScaler, and RobustScaler, which involve relatively simpler computations, where optimization space is limited, we do not exhibit significant speedup compared to CMLCompiler, but the performance difference remains within 5\%.
For more complex algorithms like LogisticsRegression, SGDClassifier, and ExtraTreeClassifier, where a larger optimization space exists, we showcase noticeable speedup due to the memory-related optimizations.

Our work achieves the best performance in 2 out of 5 tested DA operators.
We achieve a 1.04x to 5.06x speedup compared with MonetDB across 4 operators, with an average speedup of 2.38. 
We achieve a 3.34x to 108.37x speedup compared with Spark across 5 operators, with an average speedup of 28.31. 
We achieve a 1.46x to 1.58x speedup compared with pandas across 2 operators, with an average speedup of 1.1. 
For the projection operator, we adjust the data layout, thus exhibiting weaker performance compared to pandas.
However, this implementation ensures data locality, guaranteeing the performance of other operators. 
For inner join and group by operators, we showcase a 7\% and 20\% performance difference, respectively, which remains within an acceptable range. 
Overall, we achieve a noticeable speedup compared to MonetDB and Spark, and has its strengths compared to pandas.

\begin{figure*}
    \centering
    \includegraphics[width=0.98\textwidth]{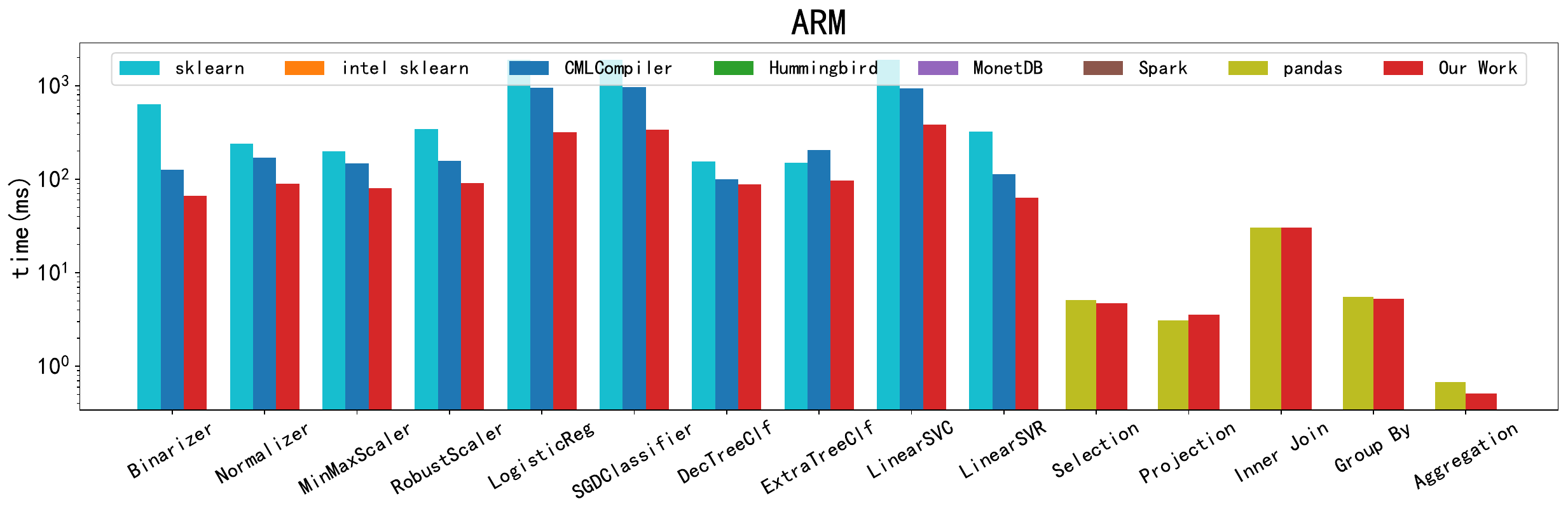}
    \caption{The performance on ARM IoT device. The absent data means unsupported.
    }
    \label{fig_arm}
\end{figure*}

Fig.~\ref{fig_arm} illustrates the performance of ARM IoT devices.
Our work outperforms across all 10 tested CML models.
We achieve a 1.56x to 9.6x speedup compared with sklearn, with an average speedup of 4.33. 
Since sklearn's high-performance library lacks optimizations for ARM IoT devices, our work obtains a larger speedup on this device. 
We achieve a 1.13x to 2.98x speedup compared with CMLCompiler, with an average speedup of 2.1. 
Intel sklearn and Hummingbird do not support this device, their results are absent. 

Our work achieves the best performance in 4 out of 5 tested DA operators.
We achieve a 1.04x to 1.32x speedup compared with pandas across 4 operators, with an average speedup of 1.06.
Except for the projection operator, which sacrifices some performance to ensure data locality, our work performs equally or better than pandas for the other operators.
MonetDB and Spark do not support this device, their results are absent. 

\begin{figure*}
    \centering
    \includegraphics[width=0.98\textwidth]{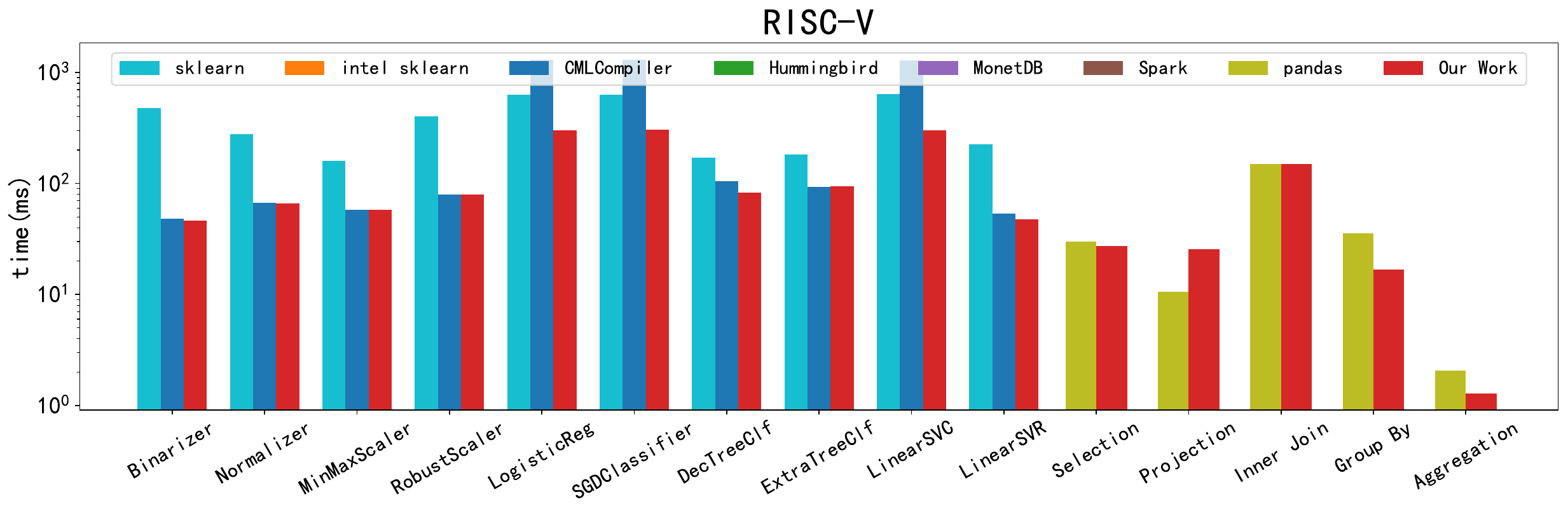}
    \caption{The performance on RISC-V IoT device. The absent data means unsupported.
    }
    \label{fig_riscv}
\end{figure*}

Fig.~\ref{fig_riscv} depicts the performance of RISC-V IoT devices.
Our work outperforms across all 10 tested CML models.
We achieve a 1.93x to 10.23x speedup compared with sklearn, with an average speedup of 3.72.
We achieve a 1.02x to 4.32x speedup compared with CMLCompiler, with an average speedup of 2.03. 
Because TVM lacks optimization for RISC-V, CMLCompiler, reliant on TVM for multi-hardware support, exhibits poor performance on RISC-V, with some algorithms even slower than sklearn. 
However, our work uses the code generator to optimize for RISC-V, resulting in a noticeable speedup compared to CMLCompiler.
Intel sklearn and Hummingbird do not support this device, their results are absent. 

Our work achieves the best performance in 4 out of 5 tested DA operators.
We achieve a 1.09x to 2.14x speedup compared with pandas across 4 operators, with an average speedup of 1.25.
MonetDB and Spark do not support this device, their results are absent.

\begin{figure}
    \centering
    \includegraphics[width=0.49\textwidth]{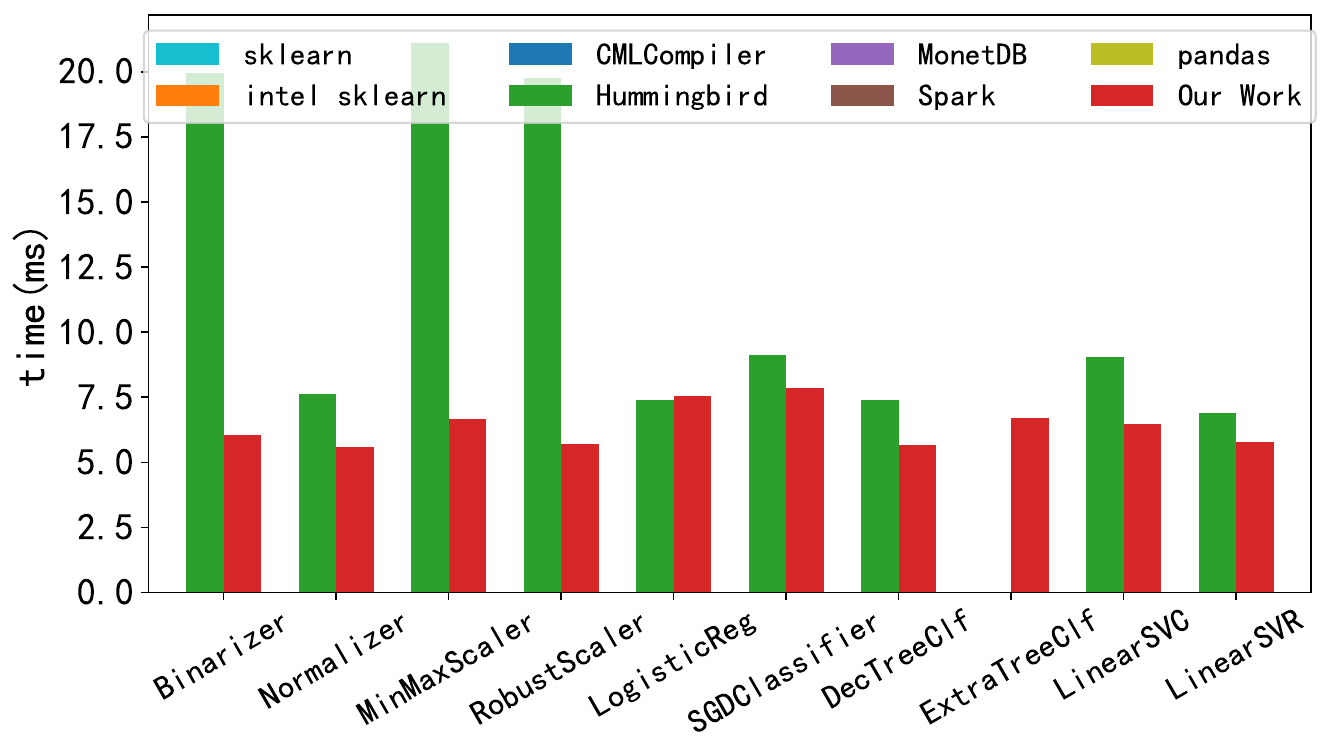}
    \caption{The performance on GPU. The absent data means unsupported.
    }
    \label{fig_gpu}
\end{figure}

Fig.~\ref{fig_gpu} shows the performance on GPU.
Our work achieves the best performance in 9 out of 10 tested CML models.
We achieve a 1.16x to 3.47x speedup compared with Hummingbird across 8 models, with an average speedup of 1.93. 
All other frameworks cannot support GPU.

In summary, compared to state-of-the-practice and state-of-the-art frameworks, our work supports a broader range of hardware and achieves noticeable speedup on them, especially for devices lacking mature high-performance libraries.
\section{Related Work}\label{sec_related_work}
This section summarizes some typical related works based on their supported application domains, portability, performance, and methods.

Deep learning (DL) frameworks include TensorFlow~\cite{10.5555/3026877.3026899}, PyTorch~\cite{NEURIPS2019_bdbca288}, MXNet~\cite{chen2015mxnet}, and Keras~\cite{gulli2017deep}. 
They are specifically designed for DL and leverage unified abstractions to make full use of domain-specific features and hardware, resulting in good performance. 
These frameworks rely on high-performance libraries such as Eigen~\cite{eigenweb}, BLAS~\cite{blackford2002updated}, and cuDNN~\cite{chetlur2014cudnn} to support hardware platforms like X86, ARM, and GPUs.
However, they are unable to support hardware platforms that do not provide the required libraries.

DL compilers include XLA~\cite{xla}, Glow~\cite{rotem2018glow}, Tensor Comprehension~\cite{vasilache2018tensor}, and TVM~\cite{10.5555/3291168.3291211}.
Similar to DL frameworks, they focus on DL and ensure performance through multi-layer abstractions.
They integrate with general-purpose compilers like LLVM~\cite{lattner2004llvm}, using code generation to support a wider range of hardware platforms, resulting in better portability.

Classical machine learning (CML) libraries include LibLinear~\cite{fan2008liblinear}, SVMlight~\cite{joachims1999svmlight}, and LibSVM~\cite{chang2011libsvm}. 
They are dedicated to several CML models and achieve good performance by implementing and optimizing algorithms from scratch. However, their portability is limited.

CML frameworks include Scikit-learn~\cite{10.5555/1953048.2078195}, Spark MLlib~\cite{10.5555/2946645.2946679}, RapidMiner~\cite{hofmann2016rapidminer}, and Weka3~\cite{frank2006weka}. 
They implement and optimize algorithms at a higher level case-by-case while leveraging high-performance libraries to support different hardware platforms. 
Their portability and performance have significant potential for improvement.

Data analysis (DA) frameworks include pandas~\cite{mckinney2011pandas}, Dask~\cite{rocklin2015dask}, Hadoop~\cite{white2012hadoop}, and Spark~\cite{zaharia2010spark}. 
They have domain-specific abstractions, leveraging high-performance libraries to support different hardware platforms. 
These frameworks exhibit good performance with moderate portability.

There are also efforts to support multiple application domains through two main approaches: extending existing frameworks and building from scratch for specific hardware. 
Databricks~\cite{dai2019bigdl} attempts to support both CML and DA in a DA framework, but it directly reuses existing abstractions, which are not suitable for DL's backpropagation and parameter iteration processes, leading to redundant computations and additional overhead that severely affect performance. 
Raven~\cite{park2022end} supports CML models in a DA framework, while H2O~\cite{h2o_platform}, Weka3~\cite{lang2019wekadeeplearning4j}, and KNIME~\cite{fillbrunn2017knime} support DL models in CML frameworks. 
Hummingbird~\cite{nakandala2020tensor} supports CML models on DL frameworks, but it lacks domain-specific optimizations.
%CMLCompiler~\cite{wen2023cmlcompiler} supports CML models on DL compilers, but it lacks optimizations for many emerging hardware such as RISC-V.
They all rely on reusing existing abstractions and invoking high-performance libraries, but cannot guarantee domain-specific optimizations, resulting in performance deviations.
NVIDIA supports all three application domains on GPUs~\cite{chetlur2014cudnn,cuml,cudf}. 
They achieve good performance by implementing and optimizing each domain's functionalities from scratch but lack portability to other hardware platforms.

%In this work, we propose a method that combines multiple layers of abstraction with domain-specific abstractions to 
Our work simultaneously supports all three application domains and provides support for a wider range of hardware platforms, ensuring both portability and performance.
\section{Conclusion}\label{sec_conclusion}
This paper proposes a methodology for constructing a unified framework to bridge the gap between domain-specific frameworks and diverse hardware devices.
The methodology leverages multi-layer abstractions, allowing applications from different domains to be represented using specific abstractions tailored to their respective fields. These abstractions are subsequently transformed into a unified abstraction, which is further translated into a minimum combination of primitive operators. Ultimately, these operators finally mapped to multiple hardware platforms. This methodology reduces the theoretical complexity of porting applications from  O(M$\times$N) to O(M + N).
A unified framework is implemented, supporting three application domains, including deep learning, classical machine learning, and data analysis on various hardware devices such as X86, ARM, RISC-V, IoT devices, and GPU.
Compared to existing solutions such as scikit-learn, hummingbird, Spark, and pandas, it supports a broader range of hardware and achieves a speedup of 1.1x to 3.83x on X86 servers, 1.06x to 4.33x on ARM IoT devices, 1.25x to 3.72x on RISC-V IoT devices, and 1.93x on GPU.

\bibliographystyle{plain}
\bibliography{references}

\begin{thebibliography}{10}

\bibitem{armcomputelib}
Arm compute library.
\newblock https://www.arm.com/technologies/compute-library, 2023.

\bibitem{cublas}
cublas: Basic linear algebra on nvidia gpus.
\newblock https://developer.nvidia.com/cublas, 2023.

\bibitem{cutensor}
cutensor: Tensor linear algebra on nvidia gpus.
\newblock https://developer.nvidia.com/cutensor, 2023.

\bibitem{xla}
Xla: Optimizing compiler for machine learning.
\newblock https://www.tensorflow.org/xla, 2023.

\bibitem{10.5555/3026877.3026899}
Mart\'{\i}n Abadi, Paul Barham, Jianmin Chen, Zhifeng Chen, Andy Davis, Jeffrey
  Dean, Matthieu Devin, Sanjay Ghemawat, Geoffrey Irving, Michael Isard,
  Manjunath Kudlur, Josh Levenberg, Rajat Monga, Sherry Moore, Derek~G. Murray,
  Benoit Steiner, Paul Tucker, Vijay Vasudevan, Pete Warden, Martin Wicke, Yuan
  Yu, and Xiaoqiang Zheng.
\newblock Tensorflow: A system for large-scale machine learning.
\newblock In {\em Proceedings of the 12th USENIX Conference on Operating
  Systems Design and Implementation}, OSDI'16, page 265–283, USA, 2016.
  USENIX Association.

\bibitem{10.1145/390013.808479}
Frances~E. Allen.
\newblock Control flow analysis.
\newblock 5(7):1–19, jul 1970.

\bibitem{backus1957fortran}
John~W Backus, Robert~J Beeber, Sheldon Best, Richard Goldberg, Lois~M Haibt,
  Harlan~L Herrick, Robert~A Nelson, David Sayre, Peter~B Sheridan, Harold
  Stern, et~al.
\newblock The fortran automatic coding system.
\newblock In {\em Papers presented at the February 26-28, 1957, western joint
  computer conference: Techniques for reliability}, pages 188--198, 1957.

\bibitem{blackford2002updated}
L~Susan Blackford, Antoine Petitet, Roldan Pozo, Karin Remington, R~Clint
  Whaley, James Demmel, Jack Dongarra, Iain Duff, Sven Hammarling, Greg Henry,
  et~al.
\newblock An updated set of basic linear algebra subprograms (blas).
\newblock {\em ACM Transactions on Mathematical Software}, 28(2):135--151,
  2002.

\bibitem{booshehri2013improving}
Meisam Booshehri, Abbas Malekpour, and Peter Luksch.
\newblock An improving method for loop unrolling.
\newblock {\em arXiv preprint arXiv:1308.0698}, 2013.

\bibitem{booth1947coding}
Andrew~Donald Booth and Kathleen~HV Britten.
\newblock Coding for arc.
\newblock 1947.

\bibitem{brosgol1980tcolada}
Benjamin~M Brosgol.
\newblock Tcolada and the" middle end" of the pqcc ada compiler.
\newblock {\em ACM SIGPLAN Notices}, 15(11):101--112, 1980.

\bibitem{callahan1988compiling}
David Callahan and Ken Kennedy.
\newblock Compiling programs for distributed-memory multiprocessors.
\newblock {\em The Journal of Supercomputing}, 2:151--169, 1988.

\bibitem{cattell1979code}
Roderic~GG Cattell, Joseph~M Newcomer, and Bruce~W Leverett.
\newblock Code generation in a machine-independent compiler.
\newblock {\em ACM SIGPLAN Notices}, 14(8):65--75, 1979.

\bibitem{chang2011libsvm}
Chih-Chung Chang and Chih-Jen Lin.
\newblock Libsvm: a library for support vector machines.
\newblock {\em ACM transactions on intelligent systems and technology (TIST)},
  2(3):1--27, 2011.

\bibitem{chen2016xgboost}
Tianqi Chen and Carlos Guestrin.
\newblock Xgboost: A scalable tree boosting system.
\newblock In {\em Proceedings of the 22nd acm sigkdd international conference
  on knowledge discovery and data mining}, pages 785--794, 2016.

\bibitem{chen2015mxnet}
Tianqi Chen, Mu~Li, Yutian Li, Min Lin, Naiyan Wang, Minjie Wang, Tianjun Xiao,
  Bing Xu, Chiyuan Zhang, and Zheng Zhang.
\newblock Mxnet: A flexible and efficient machine learning library for
  heterogeneous distributed systems.
\newblock {\em arXiv preprint arXiv:1512.01274}, 2015.

\bibitem{10.5555/3291168.3291211}
Tianqi Chen, Thierry Moreau, Ziheng Jiang, Lianmin Zheng, Eddie Yan, Meghan
  Cowan, Haichen Shen, Leyuan Wang, Yuwei Hu, Luis Ceze, Carlos Guestrin, and
  Arvind Krishnamurthy.
\newblock Tvm: An automated end-to-end optimizing compiler for deep learning.
\newblock In {\em Proceedings of the 13th USENIX Conference on Operating
  Systems Design and Implementation}, OSDI'18, page 579–594, USA, 2018.
  USENIX Association.

\bibitem{chen2019eyeriss}
Yu-Hsin Chen, Tien-Ju Yang, Joel Emer, and Vivienne Sze.
\newblock Eyeriss v2: A flexible accelerator for emerging deep neural networks
  on mobile devices.
\newblock {\em IEEE Journal on Emerging and Selected Topics in Circuits and
  Systems}, 9(2):292--308, 2019.

\bibitem{chetlur2014cudnn}
Sharan Chetlur, Cliff Woolley, Philippe Vandermersch, Jonathan Cohen, John
  Tran, Bryan Catanzaro, and Evan Shelhamer.
\newblock cudnn: Efficient primitives for deep learning.
\newblock {\em arXiv preprint arXiv:1410.0759}, 2014.

\bibitem{cocke1970global}
John Cocke.
\newblock Global common subexpression elimination.
\newblock In {\em Proceedings of a symposium on Compiler optimization}, pages
  20--24, 1970.

\bibitem{dai2019bigdl}
Jason~Jinquan Dai, Yiheng Wang, Xin Qiu, Ding Ding, Yao Zhang, Yanzhang Wang,
  Xianyan Jia, Cherry~Li Zhang, Yan Wan, Zhichao Li, et~al.
\newblock Bigdl: A distributed deep learning framework for big data.
\newblock In {\em Proceedings of the ACM Symposium on Cloud Computing}, pages
  50--60, 2019.

\bibitem{davidson1995aggressive}
Jack~W Davidson and Sanjay Jinturkar.
\newblock An aggressive approach to loop unrolling.
\newblock Technical report, Citeseer, 1995.

\bibitem{debauche2020new}
Olivier Debauche, Sa{\"\i}d Mahmoudi, Sidi~Ahmed Mahmoudi, Pierre Manneback,
  and Fr{\'e}d{\'e}ric Lebeau.
\newblock A new edge architecture for ai-iot services deployment.
\newblock {\em Procedia Computer Science}, 175:10--19, 2020.

\bibitem{Dua:2019}
Dheeru Dua and Casey Graff.
\newblock {UCI} machine learning repository, 2017.

\bibitem{eichenberger2004vectorization}
Alexandre~E Eichenberger, Peng Wu, and Kevin O'brien.
\newblock Vectorization for simd architectures with alignment constraints.
\newblock {\em Acm sigplan notices}, 39(6):82--93, 2004.

\bibitem{faghih2014land}
Ahmadreza Faghih-Imani, Naveen Eluru, Ahmed~M El-Geneidy, Michael Rabbat, and
  Usama Haq.
\newblock How land-use and urban form impact bicycle flows: evidence from the
  bicycle-sharing system (bixi) in montreal.
\newblock {\em Journal of transport geography}, 41:306--314, 2014.

\bibitem{fan2008liblinear}
Rong-En Fan, Kai-Wei Chang, Cho-Jui Hsieh, Xiang-Rui Wang, and Chih-Jen Lin.
\newblock Liblinear: A library for large linear classification.
\newblock {\em the Journal of machine Learning research}, 9:1871--1874, 2008.

\bibitem{fang2020optimizing}
Jingzhi Fang, Yanyan Shen, Yue Wang, and Lei Chen.
\newblock Optimizing dnn computation graph using graph substitutions.
\newblock {\em Proceedings of the VLDB Endowment}, 13(12):2734--2746, 2020.

\bibitem{fillbrunn2017knime}
Alexander Fillbrunn, Christian Dietz, Julianus Pfeuffer, Ren{\'e} Rahn,
  Gregory~A Landrum, and Michael~R Berthold.
\newblock Knime for reproducible cross-domain analysis of life science data.
\newblock {\em Journal of biotechnology}, 261:149--156, 2017.

\bibitem{fraboulet2001loop}
Antoine Fraboulet, Karen Kodary, and Anne Mignotte.
\newblock Loop fusion for memory space optimization.
\newblock In {\em Proceedings of the 14th international symposium on Systems
  synthesis}, pages 95--100, 2001.

\bibitem{frank2006weka}
Eibe Frank, Mark Hall, and Len Trigg.
\newblock Weka 3: Data mining software in java.
\newblock {\em The University of Waikato: Hamilton, New Zealand}, 2006.

\bibitem{eigenweb}
Ga\"{e}l Guennebaud, Beno\^{i}t Jacob, et~al.
\newblock Eigen v3.
\newblock http://eigen.tuxfamily.org, 2010.

\bibitem{gulli2017deep}
Antonio Gulli and Sujit Pal.
\newblock {\em Deep learning with Keras}.
\newblock Packt Publishing Ltd, 2017.

\bibitem{h2o_platform}
{H2O.ai}.
\newblock H2o: Scalable machine learning platform.
\newblock \url{https://github.com/h2oai/h2o-3}, 2022.

\bibitem{hofmann2016rapidminer}
Markus Hofmann and Ralf Klinkenberg.
\newblock {\em RapidMiner: Data mining use cases and business analytics
  applications}.
\newblock CRC Press, 2016.

\bibitem{huang1999generalized}
Jung-Chang Huang and Tau Leng.
\newblock Generalized loop-unrolling: a method for program speedup.
\newblock In {\em Proceedings 1999 IEEE Symposium on Application-Specific
  Systems and Software Engineering and Technology. ASSET'99 (Cat. No.
  PR00122)}, pages 244--248. IEEE, 1999.

\bibitem{jia2019taso}
Zhihao Jia, Oded Padon, James Thomas, Todd Warszawski, Matei Zaharia, and Alex
  Aiken.
\newblock Taso: optimizing deep learning computation with automatic generation
  of graph substitutions.
\newblock In {\em Proceedings of the 27th ACM Symposium on Operating Systems
  Principles}, pages 47--62, 2019.

\bibitem{joachims1999svmlight}
Thorsten Joachims.
\newblock Svmlight: Support vector machine.
\newblock {\em SVM-Light Support Vector Machine http://svmlight. joachims.
  org/, University of Dortmund}, 19(4):25, 1999.

\bibitem{karlin2013exploring}
Ian Karlin, Abhinav Bhatele, Jeff Keasler, Bradford~L Chamberlain, Jonathan
  Cohen, Zachary DeVito, Riyaz Haque, Dan Laney, Edward Luke, Felix Wang,
  et~al.
\newblock Exploring traditional and emerging parallel programming models using
  a proxy application.
\newblock In {\em 2013 IEEE 27th International Symposium on Parallel and
  Distributed Processing}, pages 919--932. IEEE, 2013.

\bibitem{1676696}
Kavi, Buckles, and Bhat.
\newblock A formal definition of data flow graph models.
\newblock {\em IEEE Transactions on Computers}, C-35(11):940--948, 1986.

\bibitem{kennedy1993maximizing}
Ken Kennedy and Kathryn~S McKinley.
\newblock Maximizing loop parallelism and improving data locality via loop
  fusion and distribution.
\newblock In {\em International Workshop on Languages and Compilers for
  Parallel Computing}, pages 301--320. Springer, 1993.

\bibitem{knoop1994partial}
Jens Knoop, Oliver R{\"u}thing, and Bernhard Steffen.
\newblock Partial dead code elimination.
\newblock {\em ACM Sigplan Notices}, 29(6):147--158, 1994.

\bibitem{lang2019wekadeeplearning4j}
Steven Lang, Felipe Bravo-Marquez, Christopher Beckham, Mark Hall, and Eibe
  Frank.
\newblock Wekadeeplearning4j: A deep learning package for weka based on
  deeplearning4j.
\newblock {\em Knowledge-Based Systems}, 178:48 -- 50, 2019.

\bibitem{lattner2004llvm}
Chris Lattner and Vikram Adve.
\newblock Llvm: A compilation framework for lifelong program analysis \&
  transformation.
\newblock In {\em International symposium on code generation and optimization,
  2004. CGO 2004.}, pages 75--86. IEEE, 2004.

\bibitem{leverett1980overview}
Bruce~W Leverett, Roderic Geoffrey~Galton Cattell, Steven~O Hobbs, Joseph~M
  Newcomer, Andrew~Henry Reiner, Bruce~R Schatz, and William~A Wulf.
\newblock An overview of the production quality compiler-compiler project.
\newblock {\em Computer}, 13(8):38--49, 1980.

\bibitem{lilja1994exploiting}
David~J Lilja.
\newblock Exploiting the parallelism available in loops.
\newblock {\em Computer}, 27(2):13--26, 1994.

\bibitem{liu2019optimizing}
Yizhi Liu, Yao Wang, Ruofei Yu, Mu~Li, Vin Sharma, and Yida Wang.
\newblock Optimizing $\{$CNN$\}$ model inference on $\{$CPUs$\}$.
\newblock In {\em 2019 USENIX Annual Technical Conference (USENIX ATC 19)},
  pages 1025--1040, 2019.

\bibitem{ma2020rammer}
Lingxiao Ma, Zhiqiang Xie, Zhi Yang, Jilong Xue, Youshan Miao, Wei Cui,
  Wenxiang Hu, Fan Yang, Lintao Zhang, and Lidong Zhou.
\newblock Rammer: Enabling holistic deep learning compiler optimizations with
  $\{$rTasks$\}$.
\newblock In {\em 14th USENIX Symposium on Operating Systems Design and
  Implementation (OSDI 20)}, pages 881--897, 2020.

\bibitem{mahlke1992effective}
Scott~A Mahlke, David~C Lin, William~Y Chen, Richard~E Hank, and Roger~A
  Bringmann.
\newblock Effective compiler support for predicated execution using the
  hyperblock.
\newblock {\em ACM SIGMICRO Newsletter}, 23(1-2):45--54, 1992.

\bibitem{mckinney2011pandas}
Wes McKinney et~al.
\newblock pandas: a foundational python library for data analysis and
  statistics.
\newblock {\em Python for high performance and scientific computing},
  14(9):1--9, 2011.

\bibitem{mehta2014revisiting}
Sanyam Mehta, Pei-Hung Lin, and Pen-Chung Yew.
\newblock Revisiting loop fusion in the polyhedral framework.
\newblock In {\em Proceedings of the 19th ACM SIGPLAN symposium on Principles
  and practice of parallel programming}, pages 233--246, 2014.

\bibitem{10.5555/2946645.2946679}
Xiangrui Meng, Joseph Bradley, Burak Yavuz, Evan Sparks, Shivaram Venkataraman,
  Davies Liu, Jeremy Freeman, DB~Tsai, Manish Amde, Sean Owen, Doris Xin,
  Reynold Xin, Michael~J. Franklin, Reza Zadeh, Matei Zaharia, and Ameet
  Talwalkar.
\newblock Mllib: Machine learning in apache spark.
\newblock {\em J. Mach. Learn. Res.}, 17(1):1235–1241, jan 2016.

\bibitem{mohammadi2018deep}
Mehdi Mohammadi, Ala Al-Fuqaha, Sameh Sorour, and Mohsen Guizani.
\newblock Deep learning for iot big data and streaming analytics: A survey.
\newblock {\em IEEE Communications Surveys \& Tutorials}, 20(4):2923--2960,
  2018.

\bibitem{nakandala2020tensor}
Supun Nakandala, Karla Saur, Gyeong-In Yu, Konstantinos Karanasos, Carlo
  Curino, Markus Weimer, and Matteo Interlandi.
\newblock A tensor compiler for unified machine learning prediction serving.
\newblock In {\em 14th $\{$USENIX$\}$ Symposium on Operating Systems Design and
  Implementation ($\{$OSDI$\}$ 20)}, pages 899--917, 2020.

\bibitem{niu2022gcd}
Wei Niu, Jiexiong Guan, Xipeng Shen, Yanzhi Wang, Gagan Agrawal, and Bin Ren.
\newblock Gcd 2: A globally optimizing compiler for mapping dnns to mobile
  dsps.
\newblock In {\em 2022 55th IEEE/ACM International Symposium on
  Microarchitecture (MICRO)}, pages 512--529. IEEE, 2022.

\bibitem{niu2021dnnfusion}
Wei Niu, Jiexiong Guan, Yanzhi Wang, Gagan Agrawal, and Bin Ren.
\newblock Dnnfusion: accelerating deep neural networks execution with advanced
  operator fusion.
\newblock In {\em Proceedings of the 42nd ACM SIGPLAN International Conference
  on Programming Language Design and Implementation}, pages 883--898, 2021.

\bibitem{nuzman2006multi}
Dorit Nuzman and Richard Henderson.
\newblock Multi-platform auto-vectorization.
\newblock In {\em International Symposium on Code Generation and Optimization
  (CGO'06)}, pages 11--pp. IEEE, 2006.

\bibitem{nuzman2006auto}
Dorit Nuzman, Ira Rosen, and Ayal Zaks.
\newblock Auto-vectorization of interleaved data for simd.
\newblock {\em ACM SIGPLAN Notices}, 41(6):132--143, 2006.

\bibitem{cuml}
{NVIDIA}.
\newblock cuml.
\newblock \url{https://docs.rapids.ai/api/cuml/stable/}, 2022.

\bibitem{cudf}
{NVIDIA}.
\newblock cuml.
\newblock \url{https://docs.rapids.ai/api/cudf/stable/}, 2022.

\bibitem{park2022end}
Kwanghyun Park, Karla Saur, Dalitso Banda, Rathijit Sen, Matteo Interlandi, and
  Konstantinos Karanasos.
\newblock End-to-end optimization of machine learning prediction queries.
\newblock In {\em Proceedings of the 2022 International Conference on
  Management of Data}, pages 587--601, 2022.

\bibitem{NEURIPS2019_bdbca288}
Adam Paszke, Sam Gross, Francisco Massa, Adam Lerer, James Bradbury, Gregory
  Chanan, Trevor Killeen, Zeming Lin, Natalia Gimelshein, Luca Antiga, Alban
  Desmaison, Andreas Kopf, Edward Yang, Zachary DeVito, Martin Raison, Alykhan
  Tejani, Sasank Chilamkurthy, Benoit Steiner, Lu~Fang, Junjie Bai, and Soumith
  Chintala.
\newblock Pytorch: An imperative style, high-performance deep learning library.
\newblock In H.~Wallach, H.~Larochelle, A.~Beygelzimer, F.~d\textquotesingle
  Alch\'{e}-Buc, E.~Fox, and R.~Garnett, editors, {\em Advances in Neural
  Information Processing Systems}, volume~32. Curran Associates, Inc., 2019.

\bibitem{10.5555/1953048.2078195}
Fabian Pedregosa, Ga\"{e}l Varoquaux, Alexandre Gramfort, Vincent Michel,
  Bertrand Thirion, Olivier Grisel, Mathieu Blondel, Peter Prettenhofer, Ron
  Weiss, Vincent Dubourg, Jake Vanderplas, Alexandre Passos, David Cournapeau,
  Matthieu Brucher, Matthieu Perrot, and \'{E}douard Duchesnay.
\newblock Scikit-learn: Machine learning in python.
\newblock {\em J. Mach. Learn. Res.}, 12(null):2825–2830, nov 2011.

\bibitem{poole1975portability}
Peter~C Poole and William~M Waite.
\newblock Portability and adaptabilty.
\newblock {\em Software Engineering: An Advanced Course}, pages 183--277, 1975.

\bibitem{potkonjak1996multiple}
Miodrag Potkonjak, Mani~B Srivastava, and Anantha~P Chandrakasan.
\newblock Multiple constant multiplications: Efficient and versatile framework
  and algorithms for exploring common subexpression elimination.
\newblock {\em IEEE Transactions on Computer-Aided Design of Integrated
  Circuits and Systems}, 15(2):151--165, 1996.

\bibitem{prakash2023cfu}
Shvetank Prakash, Tim Callahan, Joseph Bushagour, Colby Banbury, Alan~V Green,
  Pete Warden, Tim Ansell, and Vijay~Janapa Reddi.
\newblock Cfu playground: Full-stack open-source framework for tiny machine
  learning (tinyml) acceleration on fpgas.
\newblock In {\em 2023 IEEE International Symposium on Performance Analysis of
  Systems and Software (ISPASS)}, pages 157--167. IEEE, 2023.

\bibitem{ragan2013halide}
Jonathan Ragan-Kelley, Connelly Barnes, Andrew Adams, Sylvain Paris, Fr{\'e}do
  Durand, and Saman Amarasinghe.
\newblock Halide: a language and compiler for optimizing parallelism, locality,
  and recomputation in image processing pipelines.
\newblock {\em Acm Sigplan Notices}, 48(6):519--530, 2013.

\bibitem{reinders2007intel}
James Reinders.
\newblock {\em Intel threading building blocks: outfitting C++ for multi-core
  processor parallelism}.
\newblock " O'Reilly Media, Inc.", 2007.

\bibitem{rocklin2015dask}
Matthew Rocklin.
\newblock Dask: Parallel computation with blocked algorithms and task
  scheduling.
\newblock In {\em Proceedings of the 14th python in science conference}, volume
  130, page 136. SciPy Austin, TX, 2015.

\bibitem{rotem2018glow}
Nadav Rotem, Jordan Fix, Saleem Abdulrasool, Garret Catron, Summer Deng, Roman
  Dzhabarov, Nick Gibson, James Hegeman, Meghan Lele, Roman Levenstein, et~al.
\newblock Glow: Graph lowering compiler techniques for neural networks.
\newblock {\em arXiv preprint arXiv:1805.00907}, 2018.

\bibitem{rutishauser1951automatische}
Heinz Rutishauser.
\newblock {\"U}ber automatische rechenplanfertigung bei programmgesteuerten
  rechenmaschinen.
\newblock {\em Zeitschrift Angewandte Mathematik und Mechanik},
  31(8-9):255--255, 1951.

\bibitem{stallman2003using}
Richard~M Stallman et~al.
\newblock Using the gnu compiler collection.
\newblock {\em Free Software Foundation}, 4(02), 2003.

\bibitem{subhlok1993exploiting}
Jaspal Subhlok, James~M Stichnoth, David~R O'hallaron, and Thomas Gross.
\newblock Exploiting task and data parallelism on a multicomputer.
\newblock In {\em Proceedings of the fourth ACM SIGPLAN symposium on Principles
  and practice of parallel programming}, pages 13--22, 1993.

\bibitem{tan2021efficient}
Tianxiang Tan and Guohong Cao.
\newblock Efficient execution of deep neural networks on mobile devices with
  npu.
\newblock In {\em Proceedings of the 20th International Conference on
  Information Processing in Sensor Networks (Co-Located with CPS-IoT Week
  2021)}, pages 283--298, 2021.

\bibitem{vasilache2018tensor}
Nicolas Vasilache, Oleksandr Zinenko, Theodoros Theodoridis, Priya Goyal,
  Zachary DeVito, William~S Moses, Sven Verdoolaege, Andrew Adams, and Albert
  Cohen.
\newblock Tensor comprehensions: Framework-agnostic high-performance machine
  learning abstractions.
\newblock {\em arXiv preprint arXiv:1802.04730}, 2018.

\bibitem{vrevca2020accelerating}
Jure Vre{\v{c}}a, Karl~JX Sturm, Ernest Gungl, Farhad Merchant, Paolo
  Bientinesi, Rainer Leupers, and Zmago Brezo{\v{c}}nik.
\newblock Accelerating deep learning inference in constrained embedded devices
  using hardware loops and a dot product unit.
\newblock {\em IEEE Access}, 8:165913--165926, 2020.

\bibitem{wang2014intel}
Endong Wang, Qing Zhang, Bo~Shen, Guangyong Zhang, Xiaowei Lu, Qing Wu, Yajuan
  Wang, Endong Wang, Qing Zhang, Bo~Shen, et~al.
\newblock Intel math kernel library.
\newblock {\em High-Performance Computing on the Intel{\textregistered} Xeon
  Phi™: How to Fully Exploit MIC Architectures}, pages 167--188, 2014.

\bibitem{wen2024tensortable}
Xu~Wen.
\newblock Tensortable: Extending pytorch for mixed relational and linear
  algebra pipelines.
\newblock {\em BenchCouncil Transactions on Benchmarks, Standards and
  Evaluations}, 4(1):100161, 2024.

\bibitem{wen2023cmlcompiler}
Xu~Wen, Wanling Gao, Anzheng Li, Lei Wang, Zihan Jiang, and Jianfeng Zhan.
\newblock Cmlcompiler: A unified compiler for classical machine learning.
\newblock In {\em Proceedings of the 37th International Conference on
  Supercomputing}, pages 63--74, 2023.

\bibitem{werbos1990backpropagation}
Paul~J Werbos.
\newblock Backpropagation through time: what it does and how to do it.
\newblock {\em Proceedings of the IEEE}, 78(10):1550--1560, 1990.

\bibitem{white2012hadoop}
Tom White.
\newblock {\em Hadoop: The definitive guide}.
\newblock " O'Reilly Media, Inc.", 2012.

\bibitem{10.1145/321439.321440}
Maurice~V. Wilkes.
\newblock Computers then and now.
\newblock {\em J. ACM}, 15(1):1–7, jan 1968.

\bibitem{wilkes1951preparation}
Maurice~Vincent Wilkes.
\newblock {\em The Preparation of Programs for an Electronic Digital Computer:
  With special reference to the EDSAC and the Use of a Library of Subroutines}.
\newblock Addison-Wesley Press, 1951.

\bibitem{wulf1980pqcc}
William~Allan Wulf.
\newblock Pqcc: A machine-relative compiler technology.
\newblock Technical report, Carnegie-Mellon University, Department of Computer
  Science, 1980.

\bibitem{xi1999dead}
Hongwei Xi.
\newblock Dead code elimination through dependent types.
\newblock In {\em International Symposium on Practical Aspects of Declarative
  Languages}, pages 228--242. Springer, 1999.

\bibitem{xing2019dnnvm}
Yu~Xing, Shuang Liang, Lingzhi Sui, Xijie Jia, Jiantao Qiu, Xin Liu, Yushun
  Wang, Yi~Shan, and Yu~Wang.
\newblock Dnnvm: End-to-end compiler leveraging heterogeneous optimizations on
  fpga-based cnn accelerators.
\newblock {\em IEEE Transactions on Computer-Aided Design of Integrated
  Circuits and Systems}, 39(10):2668--2681, 2019.

\bibitem{yan2023hardware}
Zheyu Yan, Qing Lu, Weiwen Jiang, Lei Yang, X~Sharon Hu, Jingtong Hu, and Yiyu
  Shi.
\newblock Hardware--software co-design of deep neural architectures: From fpgas
  and asics to computing-in-memories.
\newblock In {\em Embedded Machine Learning for Cyber-Physical, IoT, and Edge
  Computing: Software Optimizations and Hardware/Software Codesign}, pages
  271--301. Springer, 2023.

\bibitem{zaharia2010spark}
Matei Zaharia, Mosharaf Chowdhury, Michael~J Franklin, Scott Shenker, and Ion
  Stoica.
\newblock Spark: cluster computing with working sets.
\newblock In {\em Proceedings of the 2nd USENIX conference on Hot topics in
  cloud computing}, 2010.

\bibitem{zhang2019openei}
Xingzhou Zhang, Yifan Wang, Sidi Lu, Liangkai Liu, Weisong Shi, et~al.
\newblock Openei: An open framework for edge intelligence.
\newblock In {\em 2019 IEEE 39th International Conference on Distributed
  Computing Systems (ICDCS)}, pages 1840--1851. IEEE, 2019.

\bibitem{zhao2022apollo}
Jie Zhao, Xiong Gao, Ruijie Xia, Zhaochuang Zhang, Deshi Chen, Lei Chen, Renwei
  Zhang, Zhen Geng, Bin Cheng, and Xuefeng Jin.
\newblock Apollo: Automatic partition-based operator fusion through layer by
  layer optimization.
\newblock {\em Proceedings of Machine Learning and Systems}, 4:1--19, 2022.

\bibitem{zheng2023chimera}
Size Zheng, Siyuan Chen, Peidi Song, Renze Chen, Xiuhong Li, Shengen Yan, Dahua
  Lin, Jingwen Leng, and Yun Liang.
\newblock Chimera: An analytical optimizing framework for effective
  compute-intensive operators fusion.
\newblock In {\em 2023 IEEE International Symposium on High-Performance
  Computer Architecture (HPCA)}, pages 1113--1126. IEEE, 2023.

\end{thebibliography}
\end{document}